\documentclass[11pt]{article}

\usepackage{graphicx}
\usepackage{amssymb}
\usepackage{hyperref}
\def\Rn{\ensuremath{\mathop{\mathrm{Re}}\nolimits}}
\def\Rc{\Rn_{\rm c}}
\def\Rg{\Rn_{\rm g}}
\def\Rt{\Rn_{\rm t}}

\title{\Large On the transition to turbulence of wall-bounded flows in general,\\
and plane Couette flow in particular}

\author{\large Paul Manneville\\
\normalsize Hydrodynamics Laboratory\\
CNRS \& \'Ecole Polytechnique, Palaiseau, France\\
\tt paul.manneville@ladhyx.polytechnique.fr}

\date{\small March 25, 2014 --- to appear in a special issue of Eur. J. Mech B/Fluids dedicated to Patrick Huerre}

\begin{document}
\sloppy
\maketitle

\begin{abstract}
The main part of this contribution to the special issue of EJM-B/Fluids dedicated to Patrick Huerre outlines the problem of the subcritical transition to turbulence in wall-bounded flows in its historical perspective with emphasis on plane Couette flow, the flow generated between counter-translating parallel planes.
Subcritical here means discontinuous and direct, with strong hysteresis.
This is due to the existence of nontrivial flow regimes between the {\it global stability\/} threshold $\Rg$, the upper bound for unconditional return to the base flow, and the {\it linear instability\/} threshold $\Rc$ characterized by unconditional departure from the base flow.

The {\it transitional range\/} around $\Rg$ is first discussed from an empirical viewpoint (\S{1}).
The recent determination of $\Rg$ for pipe flow by Avila {\it et al.} (2011) is recalled.
Plane Couette flow is next examined.
In laboratory conditions, its transitional range displays an oblique pattern made of alternately laminar and turbulent bands, up to a third threshold $\Rt$ beyond which turbulence is uniform.

Our current theoretical understanding of the problem is next reviewed~(\S{2}): linear theory and non-normal amplification of perturbations; nonlinear approaches and dynamical systems,  basin boundaries and chaotic transients in minimal flow units; spatiotemporal chaos in extended systems and the use of concepts from statistical physics,  spatiotemporal intermittency and directed percolation, large deviations and extreme values.
Two appendices present some recent personal results obtained  in plane Couette flow about patterning from numerical simulations and modeling attempts.

\end{abstract}

\paragraph{Keywords}
Transition to Turbulence; Pipe Flow; Plane Couette Flow; Laminar-Turbulent Patterning

\paragraph{Cautionary note about the literature cited} The number of articles related to the topics examined here is tremendous and, though already referring to more than 150 publications,  the bibliography is far from exhaustive by  at least one order of magnitude.
For a better coverage, the reader is invited to consult the literature cited in the review papers mentioned.
I tried not to bias the list according to my personal interests, while choosing what I thought to be the most representative papers in each subtopic, sometimes the most recent publications of given people or groups reviewing related works in their introductions and pointing backwards to earlier relevant papers.
For convenience, references are listed in alphabetical order of the first author and next chronologically.
\bigskip

\hrule
\vspace*{1.5\baselineskip}

The article published by O. Reynolds in 1883~\cite{Re83} founded the scientific approach to the problem of the transition to turbulence.
Already an abstract in itself, its title ``{\it An experimental investigation of the circumstances which determine whether the motion of water shall be direct or sinuous and the law of resistance in parallel channels\/},''  summarized the main features of the problem and, between the words, identified its control parameter $\Rn$ nowadays called the Reynolds number.
This parameter is a measure of the typical shear present in the flow under consideration.%
\footnote{Explicitly, $\Rn=U\!L /\nu$, where $U$ is a typical amplitude of velocity variations,  $L$ a typical distance over which the speed varies, and $\nu$ the kinematic viscosity of the fluid.
It can be understood as the ratio of a viscous time scale  $\tau_{\rm v}=L^2/\nu$ to an advection time scale $\tau_{\rm a}=L/U$.\label{FN1}}
When $\Rn$ is small viscous effects have enough time to tame departures from the base flow profile so that `direct motion' in Reynolds' own terms, i.e. {\it laminar\/} flow, prevails.
On the contrary, when $\Rn$ is large `sinuous motion' can be amplified up to being considered as {\it turbulent}.
The problem is then to determine/predict the value of $\Rn$ at which the transition takes place.

On general grounds two characteristic values can be defined~\cite{Jo76,MD01,HK05}, a threshold for  {\it unconditional\/} or {\it global stability\/} $\Rg$, and a threshold for {\it unconditional instability\/} $\Rc$, `c' for `critical'.
Thresholds $\Rg$ and $\Rc$ are {\it global\/} and {\it local\/} quantities, respectively.
These terms have to be understood in the general context of dynamical systems:
In the state space, `global' means whatever the amplitude and shape of the perturbation brought to the base state, whereas `local' means infinitesimal, which allows linearization and eigenmode decomposition.
$\Rc$ is obtained from linear stability analysis that can be continued in the weakly nonlinear regime around threshold by perturbation.
At this level, issues are in principle purely technical (but possibly delicate) in a well-posed setting.

Obviously, $\Rg$ lies below $\Rc$ and between $\Rg$ and $\Rc$ stability is only {\it conditional\/}: it depends on the shape and intensity of perturbations brought to the base flow.
Global stability is therefore much difficult to ascertain since the variety of possible perturbations cannot be tested in any systematic way.
In a few cases, one can show that local and global thresholds coincide, which makes the transition {\it supercritical\/}.
Thermal convection in a horizontal layer heated from below is the most celebrated example of such a circumstance~\cite{NPV77}.
This case is exceptional and, in general, permanent  departures from the base state may exist in the {\it subcritical range\/} below $\Rc$.
The {\it energy method\/}~\cite{Jo76,NPV77} generates a lower bound $\Rn_{\rm E}$ to the global stability threshold.
$\Rn_{\rm E}$ is the threshold below which the kinetic energy contained in any perturbation to the base flow decreases to zero in a monotonic way. 
This bound is usually very conservative.
By contrast, the condition defining $\Rg$ bears on the ultimate decay of the perturbations, possibly at the end of long transients during which the energy may vary wildly before decreasing like below $\Rn_{\rm E}$.

Linear instability deals with infinitesimal perturbations that can be analyzed as superpositions of elementary modes of infinite spatial extension, e.g. Fourier modes.
{\it A contrario\/}, typical perturbations living below $\Rc$ have finite amplitudes and finite supports, and coexist with laminar flow. These are the {\it flashes of turbulence\/} observed by Reynolds in his pipe or the {\it turbulent spots\/} seen in planar geometries.
When the applied shear is very large, the system is expected to be uniformly turbulent.
At least conceptually, one should therefore find another threshold separating laminar-turbulent coexistence from uniform turbulence since this represents two qualitatively different situations.
The localization of such a threshold, called $\Rt$ in the following, will also be discussed below.
What is generally called the {\it transitional range\/} is therefore the Reynolds number  interval extending from around $\Rg$ to around $\Rt$.
Table~\ref{T1} recapitulates known values of these thresholds for the two cases of main interest here, pipe flow and simple shear flow, both of them with $\Rc=\infty$.
\begin{table}
\caption{Characteristic values of the control parameter in some wall-bounded flows.
The ingredients for the Reynolds number as introduced in note~\ref{FN1}, $\Rn=U\!L/\nu$, are the mean speed  $U$ and the diameter of the pipe $D$  for Hagen--Poiseuille flow (HPF); for plane Poiseuille flow (PPF) and plane Couette flow (PCF) $L$, usually noted $h$, is the 1/2-distance between the plates; for PPF $U$ is the speed of the laminar flow in the center plane and for PCF the speed of the driving plates.  $\Rg$ is the global stability threshold,  $\Rc$ the linear stability threshold, and  $\Rt$ the threshold beyond which turbulence is featureless.\label{T1}}
\begin{center}
\begin{tabular}{  c   c   c   c  c }
\vspace*{0.4ex}
Flow & $\Rn_{\rm E}$~\cite{Jo76} & $\Rg$ & $\Rc$ & $\Rt$ \\[0.4ex]
\vspace*{0.4ex}
HPF & 81.5 & 2040~\cite{Aetal11} & $\infty$~\cite{Setal80} & $\sim 2700$~\cite{WC73} \\[0.4ex]
\vspace*{0.4ex}
PCF & 20.7 & $\sim 325$~\cite{BC98} & $\infty$~\cite{Ro73} & $\lesssim 415$~\cite{Pr01}, \S{A}  \\[0.4ex]
PPF & 49.6 & $\sim 840$~\cite{TI14} & $5772$~\cite{Or71} & $\gtrsim1600$~\cite{Ts10}\\[0.4ex]
\end{tabular}
\end{center}
\end{table}

Our understanding of  the {\it transitional range\/} in wall-bounded flows has made considerable progress recently.
Relevant information can be found in the proceedings of the 2005 IUTAM Symposium edited by Mullin \& Kerswell~\cite{MK05}.
In 2008, a whole issue of the Philosophical Transactions of the Royal Society has been devoted to the celebration of the 125th anniversary of the publication of Reynolds' article, where discussions of experimental and theoretical findings for pipe flow, also known as {\it Hagen--Poiseuille flow\/} (HPF) can be found~\cite{PTRSA367}.
Several reviews have also appeared, focussing on theoretical and numerical aspects~\cite{Eetal08,Wetal08} or on the experiments  \cite{Wetal08,Mu11}, summarizing the state of the art before 2010.
Accordingly, in \S\ref{S1.1} I shall limit myself to a brief account of posterior results in HPF centered on the quantitative determination of $\Rg$~\cite{Aetal11} that will be defined as the value of $\Rn$ below which the flashes of turbulence always decay in the long term and above which they are able to split and spread turbulence in the pipe.

With respect to simple shear flow, also called {\it plane Couette flow\/} (PCF), only partial reviews of experimental and numerical results seemingly exist~\cite{PD05,Eetal08}.
I shall not attempt to be comprehensive but try to focus on features that, in my opinion, are the most interesting.
Accordingly, in \S\ref{S1.2} I will just sketch the history of the subject and present experimental results gathered by the Saclay group~\cite{PD05} in the perspective of earlier and more recent numerical findings.
In a first series of experiments by this group, concluded with Bottin's thesis~\cite{Bo98}, the focus was on the identification of mechanisms and the determination of $\Rg$ based on the dynamics of turbulent spots in setups with moderate aspect ratio.%
\footnote{For PCF, two aspect ratios can be defined, $\Gamma_x=L_x/2h$ and $\Gamma_z=L_z/2h$ where $L_x$ and $L_z$ are the streamwise and spanwise dimensions of the shearing zone, and $2h$ the gap between the moving plates. For HPF, this would just be $L/D$ where $L$ is the length of the pipe and $D$ its diameter.\label{FN2}}
Later, in the  larger aspect ratio setup used by Prigent~\cite{Pr01}, patterns of alternately laminar and turbulent oblique bands were shown to occupy most of the transitional range, leading to the determination of the upper threshold $\Rt$.
I shall situate these findings in their context and relate them to the transition in cylindrical Couette flow (CCF) which has PCF as its small gap limit and the banded regime as the limit of {\it spiral turbulence\/} observed in that system~\cite{Co65,Aetal86,Pr01}.

Other cases of comparable interest, especially in view of applications, will not be reviewed here, in particular
{\it plane Poiseuille flow\/} (PPF), the flow between two motionless plates driven by a pressure gradient, and the Blasius boundary layer flow~\cite{SG03}.
Both are linearly unstable above some finite critical Reynolds number $\Rc$ but also display nontrivial subcritical flow in the form of turbulent spots promoting developed turbulence  when the level of residual turbulence in the base flow is large (natural transition) or when it is clean enough but appropriately triggered.
Thresholds for PPF are also quoted in table~\ref{T1}.

Some views on the present theoretical understanding of the transition will next be presented in \S\ref{S2}.
I shall first recall linear properties related to the stability of wall-bounded flows compared to free shear flows~\cite{HR98,SH01} and the importance of non-normal energy growth~Ê\cite{Tetal93,Sc07}, streamwise vortices and {\it lift-up\/}~\cite{EP75,La75}, and the process underlying the sustainment of flow patterns away from laminar flow uncovered by Waleffe and coworkers~\cite{HKW95,Wa95}.
This mechanism paves the way to sustained nontrivial states of central interest to the approach in terms of low-dimensional dynamical systems~\cite{CB13} examined in \S\ref{S2.2}.
The shift from concrete fluid mechanics to that scheme offered a conceptually appealing interpretation framework to a series of properties observed in experiments, the emergence of coexisting multiple solutions in state space {\it via\/} saddle-node bifurcations, special solutions found by refined numerical techniques explaining recurrent flow fields observed in the experiments~\cite{Hetal04}, basin boundaries and finite lifetime of turbulent flashes or spots explained through the theory of transient chaos~\cite{Eetal08}, etc.

This abstract scheme mostly relies on the concept of {\it minimal flow unit\/} (MFU), a functional setting introduced by Jim\'enez \& Moin~\cite{JM91} in which periodic boundary conditions are placed at a distance just necessary to obtain nontrivial solutions.
It implies artificial and strong confinement effects supporting the idea that the dynamics can be reduced to the strictly {\it temporal\/} evolution of well defined coherent structures.
Such constraints are however not relevant to open flow experiments in the laboratory.
In \S\ref{S2.3} I discuss the adaptations needed in view of a proper treatment of extended systems where coherence in {\it physical space\/} can no longer be taken for granted so that coexistence has now a {\it spatiotemporal\/} flavor, with regions occupied by either laminar flow or turbulence, separated by steep interfaces.
As stressed by Pomeau~\cite{Po86}, in that case a novel transition scenario can develop, {\it spatiotemporal intermittency\/}, with some features of a stochastic process called {\it directed percolation\/} that helps one to account for phenomena such as forest fires or the imbibition of porous media.
This conjecture had the virtue of  bringing the tools of statistical physics, in particular those of phase transitions and critical phenomena~\cite{St88}, into turbulence theory in a novel way.
Whereas second-order transitions with a continuous variation of the order parameter seem relevant to HPF near $\Rg$~\cite{Ba11},  PCF displays properties more in line with first-order transitions, especially through discontinuous changes associated to large deviations and nucleation~\cite{BC98,Ma09}. Recent personal simulation results related to the characterization of the transition in PCF are presented in Appendix~A, supporting the first-order schema.
Up to now, to my knowledge, patterning in PCF has not yet received any physical explanation.
Such an explanation could be obtained from well-conceived models such as the reaction-diffusion system used by Barkley~\cite{Ba11} to interpret the transition {\it via\/} spatiotemporal intermittency at $\Rg$ in HPF. Appendix~B summarizes my previous work in this domain~\cite{LM07a,Ma12} and the perspectives opened by its extension.   

\section{Brief review of the phenomenology\label{S1}}
\subsection{Global stability threshold of pipe Poiseuille flow \label{S1.1}}

Reynolds reported the natural transition in HPF to take place at $\Rn$ of the order of $2000$ via intermittent {\it flashes\/} of irregular motion with some internal structure but noted that a clean flow could be maintained laminar up to much higher $\Rn$.
Studying the turbulent flashes seen by Reynolds, Lindgren~\cite{Li57} measured the speeds of their leading and trailing edges  and called {\it slugs\/} the invasive localized turbulent domains that appear when $\Rn$ increases. (Consult also the review by Coles~\cite{Co62}.)
Later, Wygnanski and coworkers~\cite{WC73,Wetal75} scrutinized the structure of the turbulent flashes and,  in the lower part of the transitional range, identified sustained localized structures they named {\it puffs\/}.
These structures had roughly constant length of order 20 diameters, were filled with large scale mildly disordered motion, bounded upstream by a steep laminar-turbulent interface and downstream by fuzzy streaky eddies, traveling at a speed slightly smaller than the mean velocity $U$.
Upon further increasing $\Rn$ puffs were replaced by slugs filled with smaller scale turbulence and limited by laminar-turbulent interfaces of comparable steepness upstream and downstream, the leading-edge front moving faster and faster, the trailing-edge slower and slower, respectively approaching $1.6\, U$ and $0.5\, U$ at $\Rn\sim10^4$, $U$ being the average fluid speed and $2U$ the fluid speed on the centerline when the flow is laminar.

In early studies, puffs were not followed for long enough durations to decide whether they were sustained or, on the contrary, finite-lifetime transient structures, which is important in view of the determination of $\Rg$ since ultimate decay means $\Rn<\Rg$.
A novel turn was taken around 1995 when Mullin and coworkers~\cite{DM95} showed that the puffs, initially thought to be {\it equilibrium\/} states~\cite{Wetal75}, in fact probabilistically decayed to laminar flow with exponentially decreasing distributions of lifetimes.
Experiments at constant mass flux then showed that turbulence could not be sustained below $\Rn\simeq1760$~\cite{PM06,Wetal08}, proposing a first estimate for $\Rg$, with a divergence of the mean lifetime as $1/(\Rg-\Rn)$ below threshold.
Other experiments performed at constant pressure gradient (in conditions such that flow rate fluctuations were thought negligible) yielded mean lifetimes rapidly increasing with $\Rn$, faster than exponentially~\cite{Hetal06,Hetal08,Aetal10}, but without indication of singular behavior at finite $\Rn$ suggesting that  turbulence might never be sustained but only transient.

The latter claim however relied on the proviso that the dynamics of puffs does not change, which is not the case since, before being replaced by slugs, puffs see their equilibrium lengths  statistically increase with $\Rn$.
Simultaneously, they become prone to {\it splitting\/}, as early observed by Lindgren~\cite{Li57} and quantitatively studied more recently~\cite{Netal08,Aetal11,Setal14}.
Puff splitting is a complementary process that displays exponential distributions of waiting times at a given $\Rn$, with the mean waiting time rapidly decreasing as $\Rn$ increases~\cite{Aetal11}.
Whereas puff decay destroys turbulence, puff splitting spreads it further.
The two processes are therefore in competition and turbulence is statistically sustained if a puff splits before it has enough time to decay, which will inevitably happen for $\Rn$ large enough.
The winner of the competition, decay {\it vs.} splitting, changes at $\Rn=2040$~\cite{Aetal11}, which can be retained as the experimental value of $\Rg$ for HPF.
This value is reported in table~\ref{T1} since it is more in line with what I expect to happen at the global stability threshold in view of the spatiotemporal interpretation of Barkley~\cite{Ba11} further discussed in \S\ref{S2.3.2}.
By contrast, the former estimate and associated divergence would rather be the signature of a {\it crisis\/} converting a chaotic repeller into a strange attractor, within the chaotic transient paradigm of dynamical systems theory, see \S\ref{S2.2.4}. 
The reason why experiments at constant mass flux or constant pressure gradient should yield different outcomes has not yet been elucidated.

At any rate, the splitting frequency increases with $\Rn$ and splittings occur repeatedly for $\Rn> 2100$.
The region occupied by the train of daughters of a given puff expands slowly downstream and more rapidly upstream. 
Next, slugs replace plugs~\cite{WC73,Netal08,Detal10a}.
Conceptually, the value of $\Rn$  corresponding to the puff-slug transformation should be identified with the upper threshold $\Rn_{\rm t}$ since beyond, the strongly invasive character of slugs is expected to produce uniform turbulence asymptotically occupying the full pipe.
By decreasing $\Rn$ from larger values, Moxey \& Barkley~\cite{MB10} kept essentially uniform turbulence down to $\Rn=2600$ in a computational domain of length $125$ tube diameters.
On the other hand, by increasing $\Rn$ in a domain of length 400 diameters,  Avila \& Hof~\cite{AH13} numerically found that at $\Rn=2800$ the probability of turbulence breakdown over sizable intervals for limited durations remains non-negligible.
In view of these results and taking the word `transition' in a loose sense,  I have placed the intermediate value $\Rt=2700$ for HPF in table~\ref{T1}, a value which, by the way, corresponds to the upper limit for puffs given by Wygnanski \& Champagne~\cite[Fig.2b]{WC73}. 

The dynamical systems approach to coherent structures, to be considered further in Section~\ref{S2.2}, suggests that, at a given $\Rn$, the puff state belongs to a chaotic repeller with exponentially distributed lifetimes.
This can be accepted as an interpretation framework for turbulence breakdown, either at the size of the puff to explain full decay or within a puff to account for splitting.
However the approach does not tell us how the frequency of these processes vary with $\Rn$.
Reasons why exponential or even super-exponential variations of mean lifetimes might be expected have been put forward from a different perspective in terms of large deviations of processes at a MFU or sub-MFU scale~\cite{Getal10}, and similarly for the problem of waiting times involved in splitting~\cite{Setal14}.

\subsection{Laminar-turbulent patterning in Plane Couette flow\label{S1.2}}

\begin{figure}
\begin{center}

\begin{minipage}{0.54\textwidth}
\includegraphics[width=0.9\textwidth,clip]{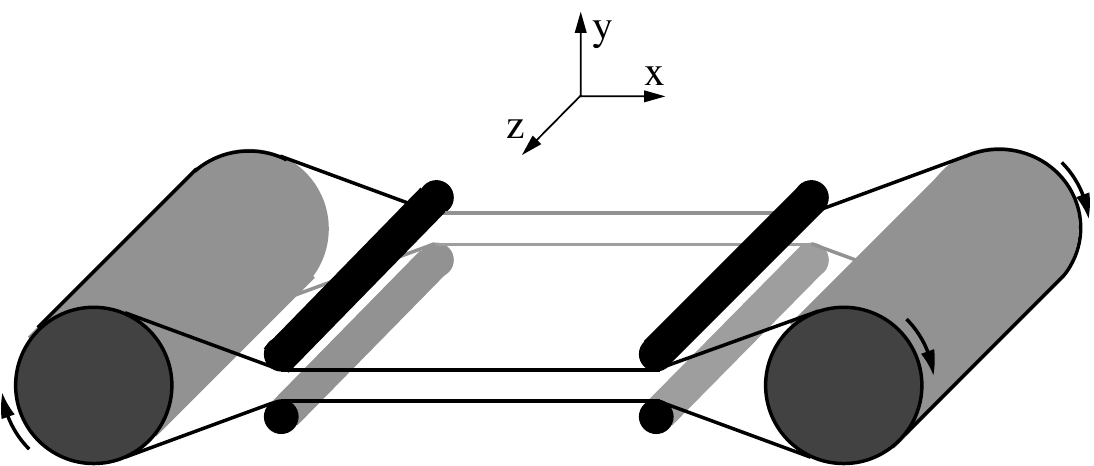}\\[2ex]
\includegraphics[width=0.35\textwidth,clip]{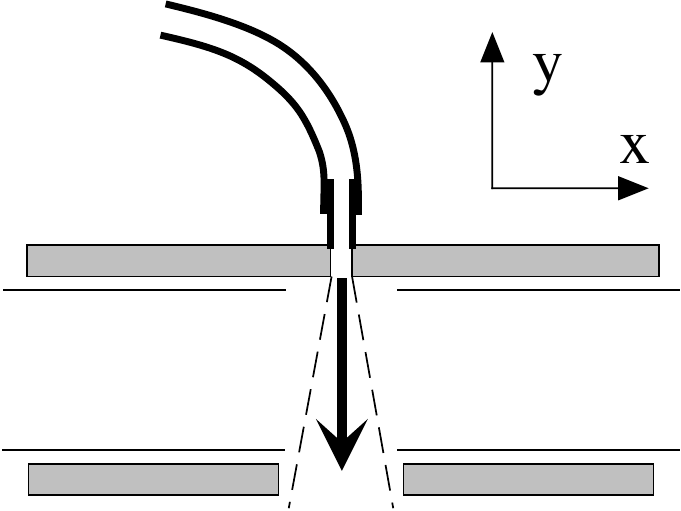}\hskip2em
\includegraphics[width=0.4\textwidth,clip]{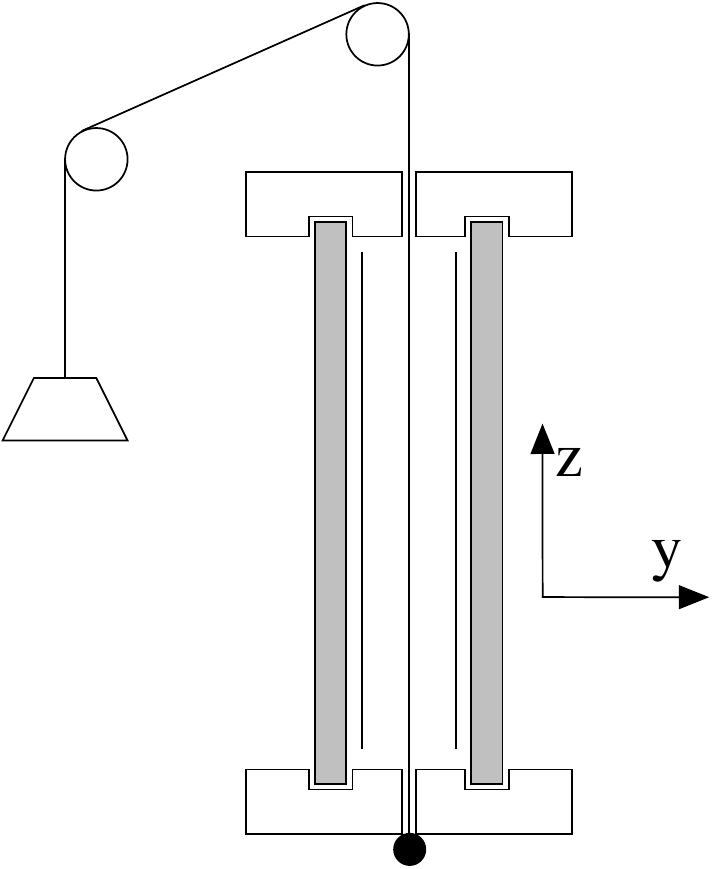}
\end{minipage}
\begin{minipage}{0.44\textwidth}
\includegraphics[width=\textwidth,clip]{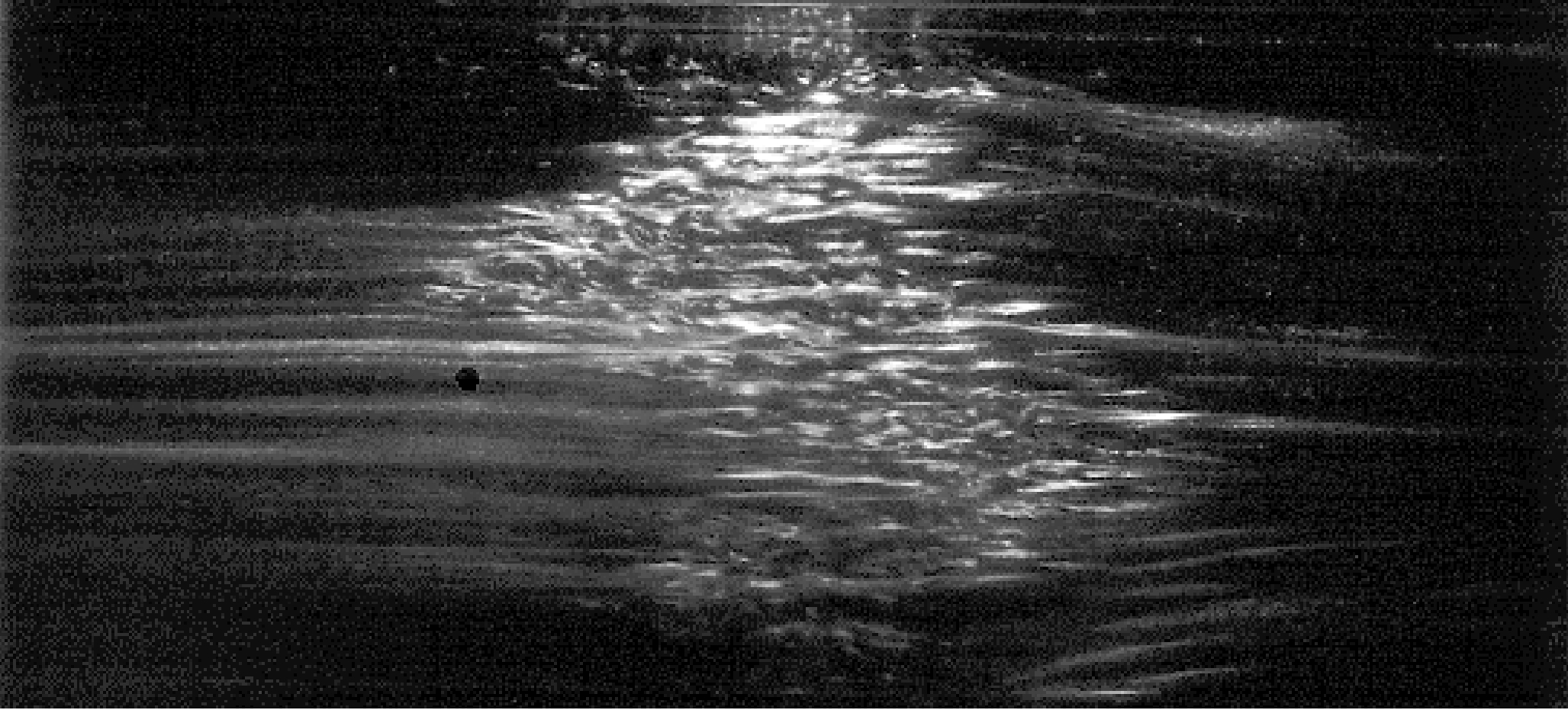}\\[2ex]
\includegraphics[width=\textwidth,clip]{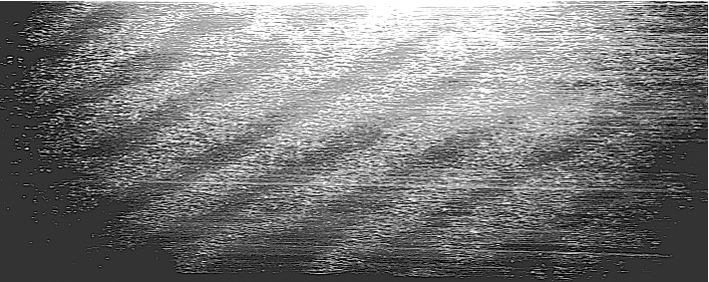}
\end{minipage}

\end{center}
\caption{\label{F1} Left: The shear is produced between the inner faces of a ribbon of width $L_z$ entrained by the big cylinders; the gap $2h$ is kept fixed by the small cylinders  \cite{TA92}; bottom: sketch of complementary setups as used at Saclay to perturb the flow: jet triggering~\cite{DD95a} and spanwise wire~\cite{DD95b}. Top-right: Mature spot at $\Rn\approx325$ in a domain $\Gamma_x\times\Gamma_z=142\times36.3$, reframed from a picture courtesy of S. Bottin~\cite{Bo98}.
Bottom-right: Oblique laminar-turbulent pattern at $\Rn=358$ in a domain $\Gamma_x\times\Gamma_z=385\times170$, after a picture courtesy of A. Prigent~\cite{Pr01}. Darker regions in the pictures correspond to laminar flow.}
\end{figure}

The other flow of interest to us, PCF, is also known to be linearly stable for all Reynolds numbers ~\cite{Ro73}.
HPF was a paradigm of a nonlinearly convectively unstable system~\cite{Ch92} in which departures from laminar flow were transported away by the mean flow.
Now, at least in experimental configurations with walls in strict anti-parallel motion (see figure~\ref{F1}),
 this difficulty is avoided and perturbations can be observed at will since they develop mostly at rest in the laboratory frame.
The difficulty has now a different origin: the geometry is basically two-dimensional in the plane of the plates, which drastically enlarges the potential number of degrees of freedom in the turbulent state, even at moderate Reynolds number.

Besides $\Rn=U h/\nu$ (note~\ref{FN1} and caption of table~\ref{T1}), aspect-ratios of the setup $\Gamma_{x,z}=L_{x,z}/2h$ (note~\ref{FN2}) turn out to be important parameters.
From the experimental point of view, the difficulty of having a very long pipe is translated into that of achieving sufficiently large aspect ratios, which puts stringent mechanical constraints since $h$ has to be small enough, hence $U$ large enough, to achieve a given $\Rn$.
References to early work can be found in the introduction of papers by  Tillmark \& Alfredsson~\cite{TA92}  and Bech {\it et al.\/}~\cite{Betal95} but a review of experimental findings equivalent to that of Mullin for HPF~\cite{Mu11} does not exist to my knowledge, which should absolve me of giving more detail.

\subsubsection{From cylindrical to plane Couette flow in the laboratory\label{S1.2.1}}

Consider first cylindrical Couette flow (CCF), the flow between differentially rotating coaxial cylinders, keeping in mind that PCF is its small gap limit.
This case is more versatile than PCF since it depends on two control parameters, the inner and outer rotation rates, besides the radius ratio.
When the inner cylinder rotates and the outer cylinder is kept at rest or rotates in the same direction as the inner cylinder, a supercritical instability develops producing Taylor vortices~\cite{Ta23,Ch81} and, after a few bifurcations~\cite{Aetal86}, turbulence is obtained at the end of a globally supercritical transition~\cite{DS85}, in close parallel to thermal convection.
When they are counter-rotating, the transition becomes subcritical and turbulent spots are observed below the linear instability threshold.
When sufficient counter-rotation is present, turbulent spots grow and arrange themselves in a continuous helical band, the {\it barber pole\/} or {\it spiral turbulence\/} regime, as observed by Coles and Van Atta long ago~\cite{Co62,Co65,VA66} and more recently by Andereck {\it et al.}~\cite{Aetal86} and others~\cite{Hetal89,LR98,GM01}.
This regime has also been reported  in numerical simulations, e.g.~\cite{Do09}.
Upon further increasing the shear, the flow enters a regime of uniform or {\it featureless\/} turbulence~\cite{Aetal86}.

In all these cases, the ratio of inner to outer radii $r_{\rm i}/r_{\rm o}=\eta$ was less than $0.9$ and, accordingly the streamwise aspect-ratio, the ratio of the mean perimeter to the gap $\Gamma_\theta=\pi(r_{\rm o}+r_{\rm i})/(r_{\rm o}-r_{\rm i})$, was kept smaller than about $60$.
In these conditions, single-branch spirals could be observed, whereas rotation and curvature effects were sizable.
The centrifugal instability of a narrow layer close to the inner cylinder~\cite{Ch81} renders the interpretation of the subcritical range more delicate.
In the counter-rotating case of interest here, this instability generates short-wavelength laminar interpenetrating spirals that become unstable.
Nonlinear focussing and breakdown of these perturbations evolve into spatiotemporally intermittent turbulent spots~\cite{Aetal86,LR98}.
As the shear rate increases, the spots grow and merge to form a complete, alternatively laminar and turbulent, {\it spiral band\/}~\cite{Co62};
next, laminar gaps close and the flow eventually enters the regime of {\it featureless\/} turbulence~\cite{Aetal86}.
In the spiral regime, the flow displays a nontrivial internal structure through the gap  at the laminar-turbulent interfaces, with laminar flow close to one cylinder facing turbulent flow near the other~\cite{Co62,VA66,Do09}, now called turbulent {\it overhangs\/}.

The dependence of the bifurcation diagram on the curvature was studied by Prigent who,  in his thesis~\cite{Pr01}, considered $\eta=0.963$ and $0.983$, hence azimuthal aspect-ratios  $167$ and $366$, respectively.
The bifurcation diagrams were found qualitatively similar with slight quantitative shifts of the different thresholds, better shown by turning to intrinsic control parameters based on mean radius and gap width as length scales, and average rotation speed and cylinder tangential speed difference as velocity scales, respectively~\cite{Na90,Ma04,PD05}.
The continuous trend from CCF to PCF as $\eta$ goes to one was particularly obvious from figure~3 in \cite{Ma04}.

With larger aspect-ratios, Prigent~\cite{Pr01,Petal02} was able to produce stable periodic arrangements of spirals with several branches and to measure the variation of the associated azimuthal and axial wavelengths with Reynolds number.
When present, the spirals were seen to rotate at the mean angular velocity and, for $\eta=0.983$, a range of rotation rates was found for which the pattern was at rest in the laboratory frame, which facilitated the comparison to the limiting case of PCF, which I now consider.

Experiments in strictly planar geometry, though conceptually simple, are scarce in the literature due to mechanical difficulties pointed out earlier.
In the {\it Fifties}, Reichardt~\cite{Re56} observed turbulence at $\Rn\sim750$ and beyond but did not considered the transitional aspects quantitatively.
At the turn of the {\it Nineties\/}, as a laboratory counterpart of early numerical experiments of turbulence growth from localized spots by Lundbladh \& Johansson~\cite{LJ91}, Tillmark \& Alfredsson~\cite{TA92}  consistently observed sustained turbulence as a result of the growth of turbulent spots around $\Rn\simeq360$.

At the same period, series of experiments were undertaken by Berg\'e's group at Saclay in order to locate thresholds either by triggering spots~\cite{Detal92,DD95a,Betal98a,BC98} (figure~\ref{F1} top-right) or by modifying the flow with tiny beads or thin spanwise wires stretched at mid-gap \cite{DD95b,Betal97,Betal98b,BC98}, either at decreasing or increasing $\Rn$. For details, consult~\cite{MD01,PD05}.
Lifetimes of turbulent spots, were shown to diverge when $\Rn$ increased toward 325~\cite{Betal98a}.
This value was also the one below which fully developed turbulence generated at large $\Rn$ was seen to decay ultimately, and also when departures from laminar flow stayed sustained in the limit of vanishing permanent perturbations (thinner and thinner spanwise wires~\cite{Betal98b}, a case also studied numerically by Barkley \& Tuckerman~\cite{BT99}).
Overall consistency of these results led to the proposal $\Rg\simeq325$ for PCF, despite a reanalysis of the variation of the mean lifetimes of spots by Hof {\it et al.}~\cite{Hetal06} suggesting a regular exponential increase rather than a singularity around $\Rn=325$.

These experiments were performed at moderate aspect ratios, $\Gamma_x\approx140$ in the streamwise direction and $\Gamma_z\approx36$ in the spanwise direction.
Halving the gap, hence doubling the aspect ratios of her current experiments, Bottin could observe obliquely arranged turbulent domains~\cite{Bo98,BC98}.
Experiments at much larger aspect ratios displaying conspicuous large scale disordered laminar-turbulent patterns~\cite[Fig.~3]{MD01} showed that this was not the end of the story.
Prigent's systematic study~\cite{Pr01,Petal02} focussed on a regular banded regime obtained by slowly decreasing $\Rn$ from high values where the flow is uniformly turbulent.
The pattern was observed%
\footnote{\label{FN4}In Prigent's manuscript~\cite{Pr01} the value $\Rt\approx415$ is indicated in figure~3.27 on p.73
but, in his figure~3.28 on p.75, the pattern is hardly visible at $\Rn=402$.
It is chevron-like at $\Rn=395$, next displays regular bands at $\Rn=376$, $358$, $349$, and disrupted bands at $\Rn=340$ and $331$. Finally, at $\Rn=323$ and $314$, the flow is fully laminar in the center of the setup but shows pattern remnants near the driving cylinders where the local Reynolds number is presumably larger than the nominal value due to entrance length phenomena. About the value of $\Rt$, see also Appendix~A.} below $\Rt\approx415$
 down to $\Rg\approx325$.
Figure~\ref{F1} (bottom-right) for $\Rn=358$ illustrates the band pattern at its strongest.
(Notice that scales are different in the two images of figure~\ref{F1}: small structures have the same size in both experiments but appear finer in the bottom image than in the top one due to the larger aspect-ratio.)

Aspect ratios much larger than previously considered were indeed necessary to observe that pattern just because its  streamwise wavelength was of the order of  $\lambda_x \simeq100$ (in units of the half-gap $h$) while, in the spanwise direction, $\lambda_z$ was regularly increasing  when decreasing $\Rn$, from about $50$ close to $\Rt$ up to $85$ in the vicinity of $\Rg$.
Comparable wavelengths were obtained in CCF at $\eta=0.983$~\cite{Pr01,PD05}, which explains that a single turbulent helical branch could be observed in setups with $\eta\le0.9$. 

\subsubsection{Numerical experiments in plane Couette flow\label{S1.2.2}}

Direct numerical simulations (DNSs) have also considerably contributed to our empirical knowledge of transitional PCF.
Many of them focussed on the determination of special solutions and will be reviewed in \S\ref{S2.2}.
Concerning the large aspect-ratios of interest to patterning, early work related either to the evolution and growth of turbulent spots~\cite{LJ91} or to the developed stage at moderate $\Rn$ but largely above $\Rt$~\cite{Ketal96}.
Fully resolved computations dedicated to the transitional range are more recent, owing to the numerical power needed.
For example Duguet {\it et al.}~\cite{Detal10b} obtained results in general agreement with laboratory experiments (thresholds, wavelengths).
Results presented in Appendix~A below, focussed on the role of periodic boundary conditions, will corroborate the experimentally observed angular variation of the bands.  

In an attempt to reduce the computational cost, Barkley \& Tuckerman \cite{BT05a,BT05b} chose to consider a long and narrow domain aligned parallel to the wavevector of the periodic modulation of turbulence characterizing the band pattern, usually at the angle with the streamwise direction expected from experiments.
So doing, they were able to reproduce most of the pattern's features.
In addition, they analyzed the structure of the mean flow inside the laminar bands, something hardly detectable in laboratory experiments, and derived from it a relation between the angle of the patterns and the Reynolds number~\cite{BT07} (for a review of their findings consult~\cite{TB11}).
By restricting the length of the computational domain to a wavelength of the pattern, in collaboration with Dauchot they determined the variation of the order parameter of the transition at $\Rt$ that they extrapolated at $\approx440$, somewhat higher than in experiments~\cite{Petal02} or fully resolved simulations in extended domains~\cite{Detal10b}.
The possibility that this shift be due to residual confinement effects in the width of their computational domain is supported by results presented in Appendix~A for domains able to accommodate at least one full pattern wavelength in the two in-plane directions.

Bands appear to be a robust feature of the transitional regime since they are appropriately rendered by decreasing the wall-normal resolution~\cite{MR11} in order to reduce the computational load differently  from Barkley \& Tuckerman.
Considering this voluntarily degraded setting as a consistent modeling strategy, I could obtain reliable hints about the local processes involved in the growth and decay of the pattern around $\Rg$~\cite{Ma11,Ma12a} to be interpreted in the theoretical discussion of~\S\ref{S2.3}.
The price to be paid was just a systematic decrease of $\Rg$ and $\Rt$, which can be understood simply by noticing that the amount of energy extracted from the base flow cannot be dissipated in sufficiently small scales so that it accumulates in the remaining degrees of freedom and keeps them chaotic at values of $\Rn$ lower than expected. 
 
Besides global subcriticality, organized laminar-turbulent alternation is certainly a characteristic feature shared by other wall-bounded flow configurations.
First, an appropriate definition of the shear Reynolds number~\cite{Ma04,Petal02,PD05} makes the range $[\Rg,\Rt]$ quantitatively correspond in the plane and cylindrical cases.
Next, at a more qualitative level, alternating laminar-turbulent banding has also recently been observed in other planar flow configurations, provided that the in-plane aspect-ratios are large enough, in numerical simulations or laboratory experiments:
in PPF (numerical~\cite{Tetal05}), in the flow between closely spaced coaxial rotating disks (torsional Couette flow, experimental~\cite{CLG02}), in the stratified Ekman layer (numerical~\cite{Deetal13}), as well as in the presence of overall rotation or other forces (numerical~\cite{Betal12,Tetal10b}), all likely a response to the same mechanism yet to be fully elucidated.
Aiming at a better analysis of flow patterns both at the local and global scale, experiments are being currently developed in PPF by Wesfreid and his group~\cite{Letal13} at ESPCI and in PCF by Couliou and Monchaux at ENSTA~\cite{Cetal14}, to mention only those who are geographically close to me.

\section{Theoretical issues\label{S2}}

\subsection{Linear {\em vs.} nonlinear approaches\label{S2.1}}

In the study of the transition from laminar to turbulent flow, the simplest case is when one can find linear instability modes serving as a first step of a cascade involving  more and more complicated dynamics.
The corresponding theory has a long story resting on linear stability analysis of parallel flows, as described, e.g., by Schmid \& Henningson~\cite{SH01}.
A cornerstone of the approach was Squire's theorem stating that most unstable modes are spanwise uniform.

Inflectional base profiles that satisfy Rayleigh's criterion are then shown to be unstable against the Kelvin--Helmholtz (KH) mechanism.
This inviscid mechanism is just mitigated by viscous effects thus responsible for instability thresholds at finite but low values of $\Rn$.
As a matter of fact, free shear layers, wakes or jets, all displaying inflection points in their respective base flow profiles, are seen to experience progressive, mostly continuous, globally supercritical transitions to turbulence at low Reynolds numbers nicely reviewed by Huerre and Rossi~\cite{HR98}.

By contrast, wall-bounded flows typically do not fulfill Rayleigh's criterion.
They can still be linearly unstable but against a subtle mechanism crucially involving viscosity.
Viscous dissipation indeed plays a counter-intuitive destabilizing role in generating Tollmien--Schlichting (TS) waves but the instability, when present, sets in at high Reynolds numbers.
The full stability problem was finally solved in the middle of the last century, when the threshold for TS waves was analytically obtained by asymptotic methods~\cite{Li55}, later confirmed numerically, e.g. for PPF~\cite{Or71}, whereas HPF or PCF were shown to be linearly stable for all Reynolds numbers~\cite{Setal80,Ro73}.
Weakly nonlinear theory {\it \`a la\/} Stuart--Landau was next developed by perturbation around spanwise-uniform neutral modes when existing.
In that way, the bifurcation of PPF at $\Rc=5772$~\cite{Or71} was shown to be subcritical and the saddle-node threshold for saturated TS waves located at $\Rn_{\rm sn}\approx2900$~\cite{He83}, still off the value $\approx1000$ where sustained and growing turbulent spots were observed~\cite{CWP82}.
The search for saturated TW waves for PCF turned out to be negative~\cite{Eetal70}, which is not surprising since this flow has no neutral modes.
 
The observation that the coherent structures filling the localized turbulent patches involved in the transition are mostly streamwise and display spanwise dependence (azimuthal in the case of HPF) suggests to forget about Squire's theorem
that does not account for algebraic perturbation growth implied by {\it lift-up\/} as pointed out by Ellingsen \& Palm~\cite{EP75} and Landahl~\cite{La75}.
The amplification of perturbation energy is due to the non-normality of the full three-dimensional linear stability operator that does not commute with its adjoint as emphasized by Trefethen {\it et al.}~\cite{Tetal93}.
Details can be found in Schmid's review~\cite{Sc07}.

As governed by the linear stability operator, mathematically infinitesimal perturbations, even strongly amplified, remain mathematically infinitesimal.
By contrast, physically small perturbations do not remain so when amplified by a large amount, thus driving the system away from its base state.
Physical ingredients for a {\it by-pass\/} scenario, e.g. oblique waves, are reviewed in~\cite{SH01}. 
In parallel, the respective role of nonlinearity and non-normality has been studied by building conceptual models in terms of simple low-order ordinary differential systems with quadratic nonlinearities preserving the energy~\cite{GG94,DM97,BT97,LM13}.
Consult Grossmann's review~\cite{Gr00} for a thorough discussion.

Beyond conceptual models, Waleffe and co-workers~\cite{HKW95,Wa95,Wa97} described a cyclic series of stages involved in the departure from base flow (below $\Rc$ for TW waves if any), commonly termed SSP for {\it self-sustaining process\/}:
At the start {\it streamwise vortices\/} are assumed.
They next generate {\it streamwise streaks\/} by lift-up.
The developing streaky flow displays a spanwise-inflectional velocity profile experiencing a KH-like instability.
The subsequent breakdown of that instability eventually regenerates the vortices.
See~\cite{Giy} for video illustrations of the SSP in PCF. 

The description of this process stemmed from a careful scrutiny by Hamilton {\it et al.} of their numerical simulations~\cite{HKW95} in domains of the size of the MFU introduced by Jim\'enez \& Moin~\cite{JM91}.
It has next been formalized using amplitudes attached to each of the modes involved in the process by means of a Galerkin method directly applied to the Navier-Stokes equations, thereby generating low-dimensional dynamical systems~\cite{Wa95,Wa97}.
The systematic use of the MFU scheme will be discussed in the next subsection devoted to the approach in terms of dynamical systems, bifurcations, and chaos theory, to which Waleffe's seminal contribution gave a large impetus.

\subsection{Transition as a dynamical system problem\label{S2.2}}

\subsubsection{Coherent structures in minimal flow units\label{S2.2.1}} 

In practice, in the systems under focus the transition to turbulence takes place at moderate Reynolds number, with coherent structures at the scale of a typical relevant length, channel height or pipe diameter.
These structures can be educed by a standard statistical analysis of the velocity correlations {\it via\/} proper orthogonal decomposition~\cite{HLB98}.
The methodology allows one to filter out smaller scales and to build empirical models governing the amplitudes of a small number of larger scale eddies.
This approach was followed by Moehlis {\it et al.}~\cite{Metal02} for PCF.

The relevance of so-obtained models relies on deep mathematical results proving that, on general grounds, the dynamics of fields governed by dissipative partial differential equations, viz. the Navier--Stokes equations, is globally attracted by low dimensional {\it inertial manifolds\/}, as described at an introductory level by Temam~\cite{Te90},
giving access to the whole framework of (finite-dimensional, dissipative) dynamical systems theory, already successfully applied to the transition to chaos in other fields of hydrodynamics~Ê\cite{SG85}, e.g. thermal convection in confined domains~\cite[Chap.~4]{Ma90}.

Objects manipulated in dynamical systems theory are phase space trajectories, limit sets (fixed points, periodic orbits, etc.) with various stability characteristics.
Manifolds attached to these elements endow the phase space with a nontrivial structure constraining the geometry of trajectories. For a recent review, consult Kawahara {\it et al.}~\cite{Ketal12}.
It should however be clear that, contrary to {\it \`a la\/} Waleffe models, these systems are never defined explicitly as sets of equations for state variables but rather implicitly from their representations in projection spaces where coordinates are observables computed from solutions to the Navier--Stokes equations.
They rest on idealizations of the physical problem, in particular by having recourse to periodic boundary conditions placed  at distances of limited relevance when compared to actual experiments.
The consequences of this change of functional setting are often under-appreciated when trying to pass to the more realistic case of extended systems because it extrapolates to large distances the strong spatial coherence enforced by periodic boundary conditions at finite (and small) distances: 
As measured in units of the half-gap $h$, typical MFU sizes for PCF are $\ell_x=7.85$, $\ell_z=4.19$, in \cite{Na90} and many subsequent studies, $\ell_x=10.89$, $\ell_z=5.46$ in \cite{Wa03}, and, at the most, $\ell_x=8\pi$, $\ell_z=2\pi$ in~\cite{Setal10b};
 for HPF, in units of the pipe diameter they are $\ell=2.1$ in \cite{FE03}, from $\ell=1.8$ to $2\pi$ in~\cite{WK04}, or $\ell=5$ in \cite{Setal07,Hetal06}.
This has however not refrained mainstream research to concentrate on the study of properties of these limit sets called {\it invariant solutions\/}.

\subsubsection{Exact nontrivial solutions to Navier--Stokes equations\label{S2.2.2}}

In the absence of linear modes allowing to obtain nonlinear solutions straightforwardly, contorted strategies are necessary to find them.
In a seminal paper, Nagata~\cite{Na90} initiated the first of two popular avenues:  {\it homotopy}, i.e. continuous deformation of the problem from a solvable case to the case of interest.
Starting from Taylor vortices in the cylindrical Couette case, he was able to find the first nontrivial steady state in PCF.
Basically the same solution was obtained later by Clever \& Busse~\cite{CB92} who started from thermal convection cells submitted to a shear.
More recently, Waleffe~\cite{Wa03} obtained a solution of the same class by continuously changing both boundary conditions and an additional bulk force to pass from a stress-free configuration to the realistic no-slip case.
For HPF, mean advection trivially transform time-independent spatially-periodic solutions into traveling waves.
Similar continuation methods allowed Faisst \& Eckhardt~\cite{FE03} or Wedin \& Kerswell~\cite{WK04} to find such traveling waves and study their properties.

These MFU solutions, once obtained can be continued as functions of $\Rn$, defining branches of a bifurcation diagram that becomes more and more complicated as supplementary modes bifurcating from them  are included, further classified according to their symmetries~\cite{Sc99,WK04}.
Nagata's solution is  time independent and bifurcates from blue sky into a pair defining two solution branches, called ``upper'' and ``lower'' as Re is increased.
The saddle-node threshold Re$_{\rm sn}$ is a function of the dimensions of the MFU but, in the cases mentioned above, is much below the experimental transition range, e.g. 125.7 in \cite{Na90} to be compared to $\approx325$.
For PCF the phase space can thus be supposed to have a basically simple structure (mostly controlled by viscosity) below that value, and a more complex nonlinear structure above~it.
Similarly, for HPF Pringle \& Kerswell~\cite{PK07} have found a pair of asymmetrical solutions down to $\Rn=773$ below which no nontrivial states seem to exist, and accordingly a simple phase space structure is expected. 

The second popular strategy, used in particular by Kawahara \& Kida~\cite{KK01} to obtain the first periodic orbits in PCF, involves performing DNSs, next identifying approximate recurrences in the record of velocity field images or observables defined from them, and eventually to refine on the trajectories by Newton iteration to achieve periodicity at a given precision level.
Two solutions were found in this way, one experiencing periodic bursting typical of the turbulent flow, the other being a gentle periodic modification of the lower branch solution evoked above.
The same methods have been applied to find special solutions in other flows, e.g. periodic states in channel flow~\cite{TI03}.
On general grounds these solutions have spatial structures that fulfill specific symmetry conditions (translations, reflections, shifts, and compositions of them), which may help understand the global structure of phase space~\cite{Getal08}.
A wealth of solutions has indeed been obtained by these methods that have been pushed to a high degree of refinement~\cite{Vi07}.
Their significance regarding the physical processes involved in the turbulence regeneration cycle is discussed in~\cite{Ketal12}.

All these solutions are saddles, i.e. unstable in a certain -- usually very small -- number of directions in phase space and attracting in the complementary subspace.
As such, they cannot be observed, but coherent structures resembling them can appear transiently as the trajectory approaches them along their attracting directions before being repelled away~\cite{Hetal04}.
The repelling directions usually correspond to perturbations that break some of their symmetries.
If the dynamics could be confined to their stable manifold -- which is not possible -- these solutions would be observable, i.e. attractors, hence the term {\it relative attractors\/}.

\subsubsection{Nontrivial states on the laminar-turbulent boundary\label{S2.2.3}}

Good examples of relative attractors are {\it edge states\/} that are special solutions sitting on the basin boundary of the base state.
This manifold separates trajectories that return to the laminar state from those that remain chaotic in the long time-limit.
Edge states generalize the concept of lower-branch state alluded to above.  
In the trivial setting of one dimensional dynamical systems, this comes to the determination of the unstable fixed point that is closest to the base state.
The approach remains straightforward in low-order conceptual models~\cite{GG94,DM97,LM13}.
Things are much more complicated for the full Navier--Stokes problem, even within the MFU framework with reduced effective dimensionality.
A working scheme has been set up first by Skufca {\it et al.}~\cite{Setal06} who determined {\it edge trajectories\/} in a shear flow model by interpolating between those that return to laminar and those that fly away at later and later times.
The same strategy has next been applied to different flows, PCF e.g.~\cite{Setal08}, HPF e.g.~\cite{Setal07}, and other flows such as the asymptotic suction boundary layer~\cite{Ketal13}.
The states obtained in that way are mildly chaotic and experience a kind of slowed-down SSP, while strongly chaotic states typical of turbulence are thought to derive from upper-branch states~\cite{Ketal12}.

Another strategy to obtain edge states and corresponding critical amplitudes has been to look for initial conditions as modes selected by non-normal amplification and subsequent nonlinear optimization.
What can be done easily in conceptual models~\cite{DM97,Co05,LM13} is much more computationally demanding for flows such as HPF~\cite{Petal12}, PCF~\cite{Detal10c,Metal11,Detal13} or  boundary layers~\cite{CdP13}, though still not free from limitations inherent in the MFU context.

The determination of edge states from the optimization viewpoint relates to an old practical question about the variation with $\Rn$ of the critical amplitude $A_{\rm c}$ of perturbations able to trigger turbulence.
Well above $\Rg$ it is expected that $A_{\rm c}$ decreases as $\Rn^{-\gamma}$ when $\Rn$ increases.
A lower bound $\gamma=1$ can be derived from a simple balance between linear and nonlinear terms in the Navier--Stokes equations but the actual value is highly debated, likely dependent on the flow geometry and a function of the by-pass scenario~\cite{Mu11}; see~\cite{WW05} or the introduction of~\cite{Detal13} for reviews.

At the price of much heavier computations, edge states can also be obtained in unbounded domains in the form of localized solutions, e.g. in HPF~\cite{Metal09}, or in PCF~\cite{Detal09}.
Flow structures resembling edge states in extended domains appear to be visited during the decay {\it from\/} turbulence in experiments~\cite{dLetal12} as well as in numerical simulations, e.g.~\cite[movie 1]{Detal10b}, or \cite{Ma11}.
By contrast, owing to their instability and the fine tuning necessary to be able to observe them during sufficiently long times in conditions of experimental relevance, such edge states are unlikely to manifest themselves in the transition {\it to\/} turbulence, by natural selection among residual fluctuations or by amplification of artificial perturbations.
For example, upon triggering HPF with the experimentally most efficient perturbations at the smallest possible amplitude, what come out  are wavy trains of  hairpin vortices that readily breakdown into puffs or slugs depending on the value $\Rn$~\cite{Tetal10}.
In the puff regime, the hairpins are easily identifiable at the start of the transient but the breakdown is so fast that the visit of the edge state, just looking like a weak puff~\cite{Metal09}, is so furtive that it cannot be pinpointed.
In view of control, the knowledge of the initial shape of nonlinearly optimal  perturbations is presumably more important than of its by-product, i.e. the edge state that will emerge from it at a later time, while the underlying chaotic dynamics leaves it trace on the discontinuous, fractal-like, dependence of the outcome on the triggering amplitude~\cite{Eetal08,Tetal10}.

\subsubsection{Chaotic transients\label{S2.2.4}}

From a mathematical standpoint, invariant solutions, especially unstable periodic orbits (UPO), play a fundamental role in chaos theory~\cite{CB13}, allowing to predict the statistics of global observables from their knowledge. 
Through the concept of {\it transient chaos\/}~\cite{Te91}, UPOs offer a nice explanation to the exponentially decaying distributions of lifetimes of puffs observed in HPF~\cite{EF05}.
As a matter of fact, the intersection of stable and unstable manifolds attached to them (support of trajectories asymptotically reaching them forward or backward in time, respectively) generate homoclinic or heteroclinic tangles rooting chaos as early envisioned by Poincar\'e~\cite{Po90}.
They form the backbone of a repeller characterized by a fractal series of lobes in which the details of the phase space trajectory seem to evolve chaotically before being expelled toward the fixed point representing the laminar flow.

It is often said that the decaying exponential distribution is the signature of a {\it memoryless process\/}.
However, when sticking to the deterministic dynamical systems point of view, on just the contrary, there is indefinite memory of initial conditions {\it which however remain unknown\/}:
Sensitivity to initial conditions, the lack of detailed knowledge about them, and the fractal structure of the repeller account for the distribution of lifetimes~\cite{Eetal08}.
It is indeed not difficult to build models producing exponentially decreasing distributions of transient lifetimes \cite[pp.240--241]{Ma90}, so that the result can be the same as for a genuinely random Poisson process but the actual explanation is conceptually very different.
 
The generic scenario describing the transition from a repeller (transient chaos) to an attractor (sustained chaos) is the {\it crisis\/}~\cite{Getal82,Ot06} taking place at some well-defined crisis condition $\Rn_{\rm cr}$ with mean lifetimes diverging as some inverse power of $|\Rn-\Rn_{\rm cr}|$.
Such a value might however not exist in wall-bounded flows within the MFU scheme.
For example, in a MFU box of size~$8\pi\times2\times2\pi$ (in units of the half-gap $h$) turbulence in PCF seems to remain in a regime of transient chaos, the mean lifetime increasing with $\Rn$ but without sign of divergence for $\Rn<\infty$~\cite{Setal10b}.
Similar results have been obtained in HPF with a short numerical pipe of length $L=5D$~\cite{Hetal06}.
This is however not a sufficient reason to exclude that, outside the restricted MFU context, there could not be a threshold beyond which turbulence is really sustained.
By placing periodic boundary conditions at finite and small distances, the MFU assumption and all the results which rests on it cannot include a whole class of perturbations able to destabilize solutions found on that basis in a quite extreme and novel way.
For PCF and other planar flows that are quasi-two-dimensional (2d) systems, periodic solutions in MFUs of in-plane size $(\ell_x\times \ell_z)$ are also solutions in domains $(n_x\ell_x\times n_z\ell_z)$ with $(n_x,n_z)$ any integers.
Forbidding perturbations of wavelengths greater than $\ell_x$ or $\ell_z$ by construction, the periodic boundary conditions strongly constrain the system and endow it with highly nonphysical spatial coherence.
For HPF, a quasi-1d system, the same  holds but only for periodicity along the axis.
This mere observation should temper the extrapolations about the transient character of turbulence~\cite{Hetal06,Setal10b}, as long as such claims are made on the basis of considerations at the MFU scale.
As a matter of fact, together with J. Philip, in~\cite{PM11} I scrutinized the passage from chaos to spatiotemporal chaos by DNS in domains of in-plane diagonal ranging from 17 ($\sim$ MFU, temporal chaos) to 140 (spatiotemporal chaos with laminar-turbulent coexistence), pointing out the  necessary decay of streamwise correlations beyond the MFU length to form the pattern observed in PCF, and calling for the introduction of standard concepts of pattern formation, as now discussed.

\subsection{Coexistence in physical space and spatiotemporal issues\label{S2.3}}

\subsubsection{Patterning {\em \`a la\/} Ginzburg--Landau\label{S2.3.1}}

The standard way to introduce space time dependence is to go from Stuart--Landau-type of equations governing the amplitude of bifurcated states to time-dependent Ginzburg--Landau (GL) equations governing the supposedly slow space-time dynamics of amplitude modulations, i.e. envelope equations~\cite{Ma90,CH93}.
The approach is most straightforward when the bifurcation is supercritical with saturating cubic nonlinearities.
For a stationary pattern forming instability like thermal convection, this yields a cubic Complex-Ginzburg--Landau equation with real coefficients, RGL3 for short.
The simplest way to pass to the subcritical case is by assuming that saturation is not achieved at lowest order but that  quintic nonlinearities will do the job, hence the RGL5 equation, with the coexistence in phase space typical of subcriticality translated into a coexistence in physical space.
Like RGL3, RGL5 derives from a potential functional.
It predicts that a domain wall between the bifurcated state and the base state moves so as to decrease the global potential, except at the value of the control parameter where the potentials corresponding to each state are equal,  the so-called Maxwell plateau well-known in the theory of first-order thermodynamic phase transitions (vis. liquid-gas).
Introducing a subdominant cubic term with a coefficient tunable {\it via\/} a non-local feedback arising from Reynolds stresses, the model put forward by Hayot \& Pomeau~\cite{HP94} offered an explanation to laminar-turbulent coexistence at scales large when compared to those involved in the SSP.
Owing to the way local and global processes were treated, the approach could however not fully account for the spiral turbulence regime of CCF, but it partly motivates the modeling of the banded regime of PCF sketched in~Appendix~B.

In the same vein, Pomeau pointed out earlier~\cite{Po86} that, when the system displays an internal periodicity scale, near the Maxwell plateau, the subcritical invasion of one state into the other becomes nontrivial due to lockings of the length of the growing domain onto multiples of the internal period.
This process was later called {\it snaking\/} by Knobloch~\cite{Kn08} from the aspect of the bifurcation diagram of corresponding multiple steady states.
For PCF developing in  domains that are spanwise extended but streamwise confined, spatial periodicity  at the scale of individual streaks may be expected to arise from the SSP.
Snaking was indeed found in that system~\cite{Setal10} but the presence of turbulence in the form of local low-dimensional chaos complicates the picture.
In that geometry, the growth of turbulent domains has indeed been shown to be more relevant to a stochastic process~\cite{Detal11} that will be discussed later.

Patterning has also been examined within the GL framework from a phenomenological standpoint, but in a completely different setting, by Prigent {\it et al.\/}  in CCF (experiments)~\cite{Pr01,Petal02} and by J. Rolland and myself in PCF (under-resolved DNSs)~\cite{RM11}.
Being interested in the transition near $\Rt$ in CCF and observing that  the spirals continuously faded away when the shear was increased, Prigent {\it et al.} considered the emergence of order at decreasing $\Rn$ as a supercritical bifurcation at increasing $(\Rt-\Rn)$.
The modulation of turbulence intensity was captured by a set of two envelopes serving as order parameters, one for each of the possible spirals, left or right.
Next, the bifurcation was described using two coupled cubic complex Ginzburg--Landau equations (CGL3) appropriate to the supercritical emergence of order.
Finally, noise was added to account for background turbulence in the featureless regime, transforming the equations into a Langevin system.
This description was then put on a quantitative basis by fitting the parameters in the model against experiments.
Spirals were shown to be well described by GL equations with all their coefficients real, supporting the claim they are not propagating waves but more simply static modulations of the turbulence intensity trivially transported by the mean rotation rate.
The variation of the  nonlinear interaction coefficient with the control parameter also appeared fully compatible with the fact that a single spiral orientation is selected far enough from $\Rt$.
Finally and perhaps not unsurprisingly, the Langevin-like contribution was shown to delay the ordering and lower $\Rt$ to an apparent value much below what could be extrapolated from the well established nonlinear regime far from $\Rt$.
Similar results were obtained in PCF~\cite{RM11}.
Tuckerman {\it et al.}~\cite{Tetal09} also characterized that transition using their narrow and oblique computational domain. They obtained $\Rt\approx440$ significantly larger than the value found in the experiment $\Rt\approx 415$ (itself probably somewhat overestimated; see note~\ref{FN4}, p.~\pageref{FN4}).
The two findings could be reconciled by recognizing that the effects of orientation fluctuations present in the laboratory~\cite{Petal02} and in the numerics in extended domains~\cite{Detal10b} are killed in the oblique but narrow computational domain:
Fixing the orientation beforehand strengthens the flow coherence and pushes the advent of  featureless turbulence to larger values of $\Rn$.
The study to be presented in Appendix~A supports this interpretation.

\subsubsection{Directed percolation and spatiotemporal intermittency\label{S2.3.2}} 

Extending his analysis of the peculiarities of front propagation in spatially unfolded subcritical cellular instabilities, Pomeau ~\cite{Po86} also suggested that, when one of the states in competition is chaotic, the dynamics of the front separating domains in different states is presumably stochastic.
He further conjectured that the process underlying the laminar-turbulent coexistence is akin to {\it directed percolation\/} (DP) to be studied using the tools of the theory of critical phenomena~\cite{St88}.

In statistical physics, DP~\cite{Ki83} is a stochastic process defined on a lattice where nodes can be in one of two states, either `on' or `off', usually termed {\it active\/} and {\it absorbing\/} respectively.
Next, a node in the active state can spontaneously decay to absorbing with some probability whereas a node in the absorbing state can only be reactivated by contamination from active neighbors, again with some probability.
Activity propagates to infinity if contamination is strong enough, otherwise the final state is entirely absorbing.
Transitional wall-bounded flows are obviously eligible to such a framework:
Local chaos features the active state with finite lifetime, while laminar flow, being linearly stable, is absorbing in the required sense.
Accordingly, turbulence should spread by contamination only, with the {\it  turbulent fraction\/} -- the relevant order parameter -- taking non zero values only above the putative threshold when contamination is able to overcome spontaneous decay at large enough Reynolds numbers.

The Ising model works as  a prototype for phase transitions in equilibrium thermodynamics~\cite{St88}.
It accounts for the ferromagnetic ordering of a paramagnet below its {\it critical temperature\/} (the Curie point).
In the same way, DP is the prototype of models for non-equilibrium phase transitions, owing to the irreversible nature of the decay into the absorbing states~\cite{Hi00}.
In both cases, the statistical properties of the system considered do not depend on its microscopic definition but on global characteristics such as the symmetries of the order parameter and the space dimension, defining {\it universality classes\/}.
Near the critical value of the control parameter, temperature for the ferromagnetic transition, decay or contamination probability for DP, the statistical properties are singular and vary as power laws of the distance to criticality characterized by  sets of {\it critical exponents\/}~\cite{St88}.

In fact, DP is a purely stochastic process whereas the description of fluid flow {\it via\/} Navier--Stokes equations is deterministic.
Deterministic processes behaving as DP can however be easily constructed in terms of coupled map lattices~\cite{CM95}.
The simplest implementation relevant to our problem involve arrays of sub-systems governed by identical maps displaying transient chaos close to a crisis point~\cite{Ot06} and coupled to neighbors on the lattice through diffusion.
A given node of the array stays active as long as its evolution is chaotic, becomes absorbing after a certain number of iterations depending sensitively on its initial condition, but can be reactivated through coupling to its neighbors if some of them are still active.
In that way, transient local chaos can be converted into sustained spatiotemporal chaos upon variation of the coupling strength or some parameter controlling the local dynamics.
{\it Spatiotemporal intermittency\/} (STI) is the term coined to refer to this kind of transition that reconciles determinism and stochasticity for systems distributed in space, stochasticity being induced by chaos in the local dynamics.
Above all, it permits us to understand that {\it turbulence can be sustained even if, at the local scale, it is only transient}. 

In one space dimension, Barkey~\cite{Ba11} constructed a two-variable coupled map lattice mimicking transitional HPF along the lines suggested above, with local dynamics apt to describe transient local chaos, completed by diffusive coupling with neighbors, and a term accounting for advection of the shear.
By adjusting a small number of parameters and varying just one featuring $\Rn$, he was able to reproduce the different regimes of puff decay, puff splitting, and continuous turbulence growth observed experimentally~\cite{Aetal11}, giving evidence that the transition at $\Rg$ was DP-like, with exponents in the corresponding universality class (at least within his model).

\subsubsection{Nucleation, large deviations and extreme values\label{S2.3.3}}

Though DP critical exponents can be recovered in some cases~\cite{Ba11,Setal13}, STI appears to be richer than plain DP, in particular with respect to the continuous/discontinuous character of the transition~\cite{Betal01}.
With an appropriate local dynamics, the STI transition can indeed be discontinuous and first-order like.
This possibility was explored by  Bottin \& Chat\'e~\cite{BC98} who showed that, in two space dimensions (the in-plane directions of PCF) the decay of sustained spatiotemporal chaos may be due to the nucleation of sufficiently wide laminar holes.

Nucleation was not explicitly mentioned by Pomeau in~\cite{Po86}, but he scrutinized the analogy with first-order transitions in a much deeper way in a follow-up of the book with Berg\'e and Vidal~\cite{BPV98}.
Written in French this second book was unfortunately never translated into English.
He insisted on the concept of {\it germ\/}, a fluctuation in a region of physical space that has tumbled from the formerly stable state into the new state, and that of {\it critical germ\/}.
In the transition {\it to\/} turbulence, artificial germs were used in triggering experiments and the amplitude of the perturbation was the control parameter.
Here, in the transition {\it from\/} turbulence, natural breakdown of featureless turbulence is of interest and fluctuations are intrinsic to the turbulent regime.
The critical quantity is rather the size of the fluctuation: below critical it recedes in the long term, above critical it grows and the new state irreversibly invades the whole system.
The model used by Bottin \& Chat\'e was just analogical and did not attempt to reproduce Navier--Stokes dynamics.
Using a reduced model derived from primitive equations~\cite{LM07a}, I studied the decay of turbulence and could show that featureless turbulence indeed breaks down when $\Rn$ is decreased similarly to what can be expected from a nucleation process~\cite{Ma09}.
The distribution of sizes of laminar holes was studied as a function of $\Rn$.
They were shown to display power-law tails in the form $\Pi(S)\sim S^{-\alpha}$ where $S$ is the surface of the laminar hole.
Exponent $\alpha$ was found to decrease from values $>3$ at high $\Rn$ to values $<3$ when the relaxation to laminar flow was observed, at the end of a long transient during which a laminar germ of size above some critical size was seen to invade the rest of the domain still in the STI state.
The value $\alpha=3$ quoted above corresponds to a threshold below which the variance of the distribution of sizes $S$ is infinite, which means that the occurrence of a germ beyond the critical size is bound to appear in the system, while for $\alpha>3$ the variance is finite so that large germs beyond critical size have a negligible probability to appear.
The main consequence of this result was that the conversion of transient local chaos into sustained global spatiotemporal chaos happens {\it whether or not the lifetime of local chaos diverges\/}, this local lifetime being obtained in the low-dimensional dynamical system setting valid in the MFU context.

The model I considered \cite{Ma09} was however imperfect in that it did not reproduce the bands (see~Appendix~B).
Studying the decay of bands for $\Rn\lessapprox\Rg$ in DNSs of Navier--Stokes equations~\cite{Ma11}, I could in fact observe a nice combination of the two processes described earlier:
{\it i\/}) the nucleation of a laminar gap of length commensurate to the width of the turbulent band itself, and {\it ii\/}) a slow stochastic withdrawal of turbulence by elementary steps corresponding to the breakdown of turbulence at the scale of a streak.
The second stage is essentially what can be expected for the retraction of an active domain below the percolation threshold in 1d (here in the direction of the length of the band fragments left after the nucleation of a laminar gap).
The complementary problem of growth from a germ for $\Rn\gtrapprox\Rg$ was also studied in the same conditions~\cite{Ma12a}.
As soon as the germ is sufficiently large it takes an oblique shape with a given orientation.
At large scale it grows by nucleating turbulent blobs of the same or the opposite orientation, near its tips.
This process stays in competition with the collapse of band fragments of similar size, but with a bias toward growth.
At the same time, a local stochastic birth-and-death process governs the streaky structures, again biased toward growth as in DP above threshold.

A proper account of the effective two-dimensionality of PCF seems essential since, by contrast, spanwise elongated but streamwise narrow computational geometries that do not permit orientation changes only produce processes closer to standard DP, either at decay~\cite{Setal13} or during growth~\cite{Detal11}.
A complete 2d treatment is however complicated by the presence of {\it large scale flows\/} that develop as a response to mean Reynolds stress distributions~\cite{HP94,BT07,LM07b}.
They contribute to all processes active in the transitional range, in particular in the growth of spots from germs as soon as they become sufficiently large, breaking the streamwise symmetry~\cite{DS13}, shaping the bands~\cite{Ma12a}, and introducing nonlocal effects possibly responsible for the selection of their wavelength~\cite{HP94} (see also Appendix~B).

Nucleation is a large deviation process that has already been proposed to account for puff collapse in HPF~\cite{Getal10}.
Studying band breaking in PCF~\cite{Ma11}, I arrived at the same conclusion that turbulence decay involves large deviations nucleating sufficient laminar gaps.
Laminar gaps, large or small, leave their mark on the average turbulent energy $E_{\rm t}$ contained in the flow as dips, deep or weak, in the time series of that quantity.
Below $\Rg$, when turbulence is transient, $E_{\rm t}$ ultimately decays to zero.
Above $\Rg$ it is not supposed to do so but it fluctuates greatly and its time series displays minima that indicate more or less pronounced failed attempts to relax to laminar flow.
In collaboration with Lucarini and his coworkers~\cite{Fetal14}, I studied the statistical properties of breakdown in the vicinity of $\Rg$ by considering the distribution of extreme minima of $E_{\rm t}$ in a systematic but strictly empirical way. 
Depending on the value of $\Rn$, the series of minima were distributed according to one of three possible extreme-value laws, with behaviors discriminated by the sign of the {\it shape parameter\/} $\kappa$ that describes the tail of the distribution~\cite{Letal83}.
At larger $\Rn$, $\kappa$ was negative, indicating a Weibull distribution that is bounded from below, hence sustained turbulent bands.
When $\Rn$ decreases, the sign of $\kappa$ changed and the distribution changed to a Fr\'echet law with exponentially decaying tail, hence exponentially small but finite probability to decay.
Even if the experiment was not long enough to observe decay, the value of $\kappa$ determined from finite (but still long enough) time series could thus tell us whether turbulence was transient or not, whether $\Rn$ was below or above $\Rg$.
In this perspective the value of $\Rn$ when  $\kappa=0$, i.e. a Gumbel distribution for the minima, defines $\Rg$ in an objective way for extended open flows of experimental interest~\cite{Fetal14}.

\section{Concluding remarks\label{S3}}

Over the years, the  {\it problem of turbulence\/} has rather been treated as the {\it problem of the transition to turbulence\/}, as formulated by Landau~\cite{La44} and Ruelle \& Takens~\cite{RT71}.
Implicit was the modal approach directly stemming from the conventional approach to hydrodynamic stability.
By {\it modes\/} it is generally meant spatiotemporally coherent structures evolving under some specific autonomous dynamics that can further serve as elementary bricks to understand the behavior of the system as a whole {\it via\/} their interactions.
Most usually these coherent structures arise from {\it neutral\/} eigenmodes of the linear stability operator further extrapolated to the weakly nonlinear regime.
This situation is typical of {\it normal systems\/}, of which Rayleigh--B\'enard convection is a good example, convection cells being the relevant coherent structures~\cite{NPV77}.
The transitional range of wall-bounded flows is not amenable to such an approach because the laminar profile is linearly stable in that range, so that there are no neutral modes to start from.

First considering the transition {\it to\/} turbulence, $\Rn$ being swept up, non-normal amplification and by-pass are the rules.
Obtaining relevant coherent structures is a hard task that has been accomplished only recently within the framework of dynamical systems theory~\cite{Eetal08,Ketal12}.
The MFU concept has been instrumental in this quest, focussing on nonlinear interactions in a context of strong coherence reinforced by the periodic boundary conditions set at short distances.
All the concepts associated with {\it temporal chaos\/} are then fully applicable but may not be relevant since crucial perturbations breaking the symmetries of the coherent structures, especially those corresponding to subharmonic modulations, are forbidden by assumption.
As a consequence, over the whole range of Reynolds numbers relevant to the transition, while local chaos is most certainly transient as long as strict MFU conditions are assumed, turbulence understood as {\it spatiotemporal chaos\/} can still be sustained due to interactions at scales larger than the MFU size:
From the theory of pattern formation~\cite{CH93,Ma90} one can expect that the effective number of degrees of freedom involved in the system under consideration is an {\it extensive\/} quantity~\cite{Ru82,PM11} that increases like its size evaluated in terms of the number of coherent structures restricted to supposedly isolated MFUs, i.e. linearly with the axial coordinate in HPF and quadratically with the streamwise and spanwise direction in PCF and other effectively two-dimensional systems.
Furthermore,  a wide class of easily excitable perturbations is now accessible, that render the coherent structures highly unstable as soon as domains a few MFU wide and long are considered, much more unstable than what can be estimated on a local basis, especially owing to the existence of long-wavelength (spatial) phase modes.

This very observation supports my belief that, in all circumstances of practical interest for wall-bounded flows, the dynamical system approach at the MFU scale is really valuable to identify nonlinear mechanisms at work, e.g. the SSP~\cite{HKW95,Wa97} and other more subtile effects like bursting~\cite{Ketal12}.
But I also think that it cannot give reliable information on the long term sustainment of turbulence, on its statistics when sustained, nor on the shape of the most dangerous perturbations living on the basin boundary of the laminar flow.
If strict coherence is no longer the master word in even moderately extended systems, tools from different fields are necessary.
I have been advocating some of those that were developed in a statistical physics framework and can give definite answers in these circumstances, e.g. the decay/growth of turbulent spots in terms of large deviations in weakly confined PCF domains, or the decay/splitting of puffs in realistic experimental conditions~\cite{Getal10,Setal14}. 

Now considering the transition {\it from\/} turbulence in extended geometries, $\Rn$ being swept down,
we do not face a conventional situation with laminar competing states, base state and bifurcated state, for which local thermodynamic equilibrium allows one to deal with a stability problem in a deterministic framework due to negligible thermal fluctuations.
Here the case is interesting since the reference state, i.e. featureless turbulence at high $\Rn$, exhibits strong macroscopic fluctuations of intrinsic origin.
Though turbulent noise is not thermal noise, it is tempting to consider the analogy with thermodynamics equilibrium and thermodynamic transitions seriously and to understand turbulence collapse as a first-order phase transition~\cite{BPV98}.
Laminar-turbulent state coexistence would then correspond to a Maxwell plateau without properly defined free energy, with turbulence possibly ``under-cooled''  and laminar flow ``over-heated.''
The actual transition would be controlled by the presence of germs bigger than some {\it critical germ\/} arising spontaneously in the turbulent phase or artificially in the laminar phase as long as it is linearly stable.

Whatever the intensity of the noise, the first step in the theoretical approach to phase transitions is the mean field approximation~\cite{St88}, rendering the problem equivalent to deterministic bifurcation theory.
It becomes therefore natural to try to extend the approach to patterning observed in the transitional range of PCF and to understand it as the result of a standard cellular instability of the featureless regime, which is the main objective of the reductionist attempt sketched in Appendix~B.
Trying to identify mechanisms for turbulence intensity modulations from the consideration of primitive equations indeed seems an interesting challenge.

To conclude, the long standing problem of the transition to turbulence has gone through important advances from laboratory and computer experiments, appropriately combined to mathematical developments  and concepts borrowed from  statistical physics.
This progress certainly improves our understanding of specific transitional flows that might benefit applications to control in other wall-bounded flows.
From a more general viewpoint, the problem of patterning over a turbulent background also questions the general theory of far-from-equilibrium nonlinear dynamics and complex systems in an original way.

\appendix

\section{Simulations of PCF in domains of moderate size\label{A1}}

\subsection{Computational methodology\label{A1.1}}

Numerical simulations of PCF have been performed using Gibson's open software {\sc ChannelFlow}~\cite{Gix}.
This pseudo-spectral code treats the wall-normal dependence $y$ using Chebyshev polynomials and the in-plane dependence (streamwise $x$, spanwise $z$) by fast Fourier transforms.
Domains able to fit at least one pattern wavelength ($\lambda_x,\lambda_z$) in each direction have been chosen.
Two sizes have been considered, a small one  $\mathcal D_{\rm s}$ with $L_x=108$ and $L_z=64$, i.e. one oblique band, and a large one $\mathcal D_{\rm b}$ with $L_x=128$ and $L_z=160$, i.e. two to three oblique bands~\cite{Pr01}.
According to previous work~\cite{Detal10b,BT05b,MR11}, the resolution is thought to be reasonably good with $N_y=33$ Chebyshev polynomials, and  $N_x=3L_x$, $N_z=6L_z$, where $N_{x,y}$ are the numbers of Fourier modes after de-aliasing (three-half the number of evolving Fourier modes).
Accordingly, no significant shift of the transitional range $[\Rg,\Rt]$ is expected~\cite{MR11}.
Some of the results below have been presented at the ETC14 conference~\cite{Ma13}.

Several experiments have been performed starting from a uniformly turbulent state at $\Rn=470$ stemming from random initial conditions.
The Reynolds number was then decreased regularly by steps $\delta\!\Rn$ down to turbulence decay.
Plateaus in $\Rn$ had a fixed duration $\delta t$.
The state obtained at the end of a plateau served as an initial condition for the next step downwards while, on the other hand, the simulation at the current value of $\Rn$ was continued till statistical equilibrium was reached and maintained steady over a significant period of time.
Steadiness was appreciated from movies and from time series of ``distance'' to laminar flow measured as the average mean perturbation energy $E_{\rm t}= [2L_xL_z]^{-1} \int_{-1}^{+1} \frac12 \mathbf v^2 \,{\rm d}x \,{\rm d}y\,{\rm d}z$ (where $\mathbf v$ is the departure from laminar flow), typically over more than $5\times10^3\>h/U$.

When the duration $\delta t$ of the plateau is small, typically $500~h/U$ or less, the system has just enough time to learn that the Reynolds number has changed -- viscous time $\tau_{\rm v}$ in units of $h/U$ is numerically equal to $\Rn$ -- but not enough time to adjust its pattern.
In order to keep the flow turbulent at values of $\Rn$ as low as possible, I chose to avoid quenching the flow from large values of $\Rn$ where the flow is uniformly turbulent down to values in the transitional range.
This is because deep quenches are immediately followed by a viscous relaxation stage that may leave an insufficient disturbance level from which turbulence has a finite probability not to recover, even if the final $\Rn$ is well above~$\Rg$.

\subsection{Results\label{A1.2}}

In the present study, I have been  primarily interested in the Fourier analysis of the large scale spatial modulation of the perturbation energy averaged over the upper layer $0\le y\le 1$, $e_{\rm t} (x,z,t) = \frac12 \int_0^1\mathbf v^2 {\rm d} y$, which captures the turbulent activity quite well.
Similar results have been obtained with $\mathcal D_{\rm s}$ and $\mathcal D_{\rm b}$ but larger confinement effects are expected in the smaller domain.
Laboratory experiments show that the angle of the pattern with the streamwise direction increases as $\Rn$ decreases.
So, the larger domain $\mathcal D_{\rm b}$ was chosen to study how this orientation change can be reproduced by numerical simulations with periodic boundary conditions. 

In both cases, as long as the turbulent regime was essentially uniform, upon decreasing $\Rn$ from 470 as described earlier with $\delta t=500$ and $\delta\!\Rn=20$ or $10$, the system rapidly equilibrated while staying featureless.
As the transitional range was approached, I still kept  $\delta t=500$ from $\Rn=320$ downwards but slowed down the decrease of $\Rn$ by taking $\delta\!\Rn=5$.

\begin{figure}
\begin{center}
\includegraphics[width=0.8\textwidth]{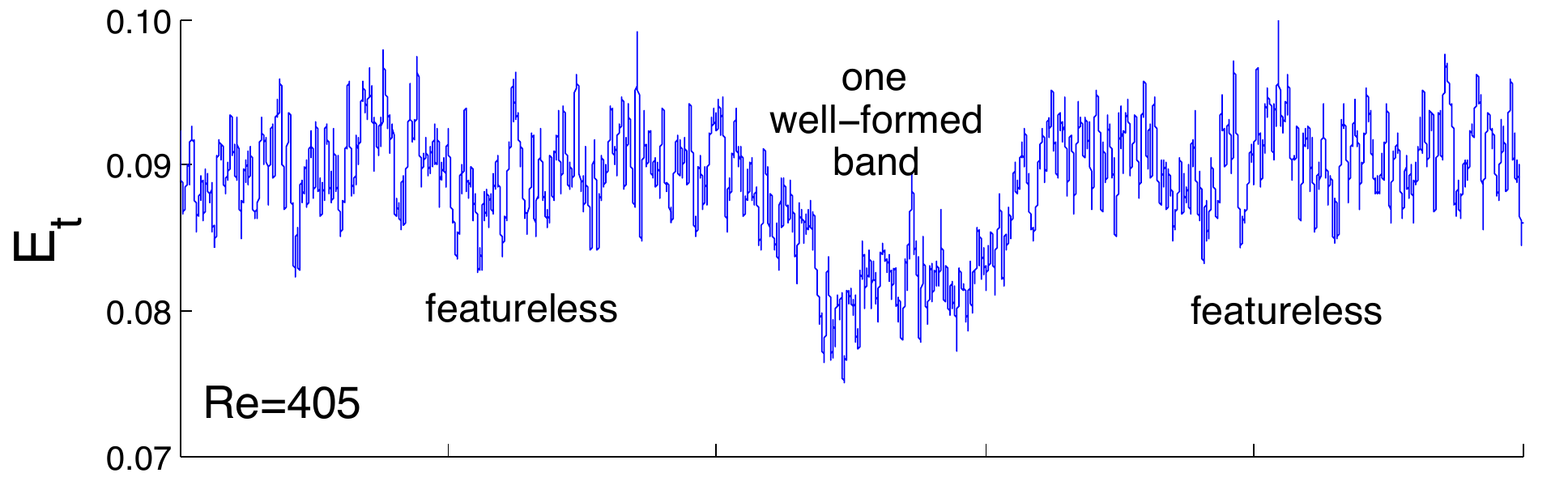}\\
\includegraphics[width=0.8\textwidth]{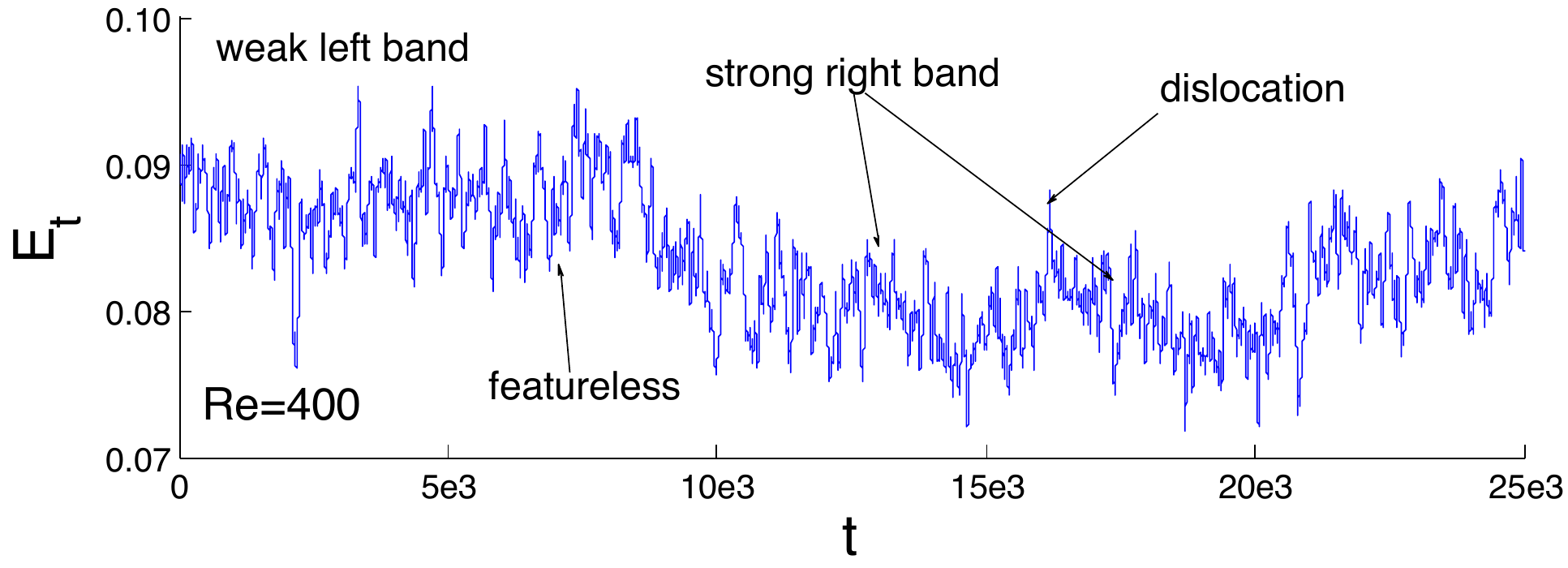}
\end{center}
\caption{\label{Fet}Time series of the turbulent energy in the small domain ($108\times64$) for $\Rn=405$ and $400$, suggesting a discontinuous transition at $\Rt$.}
\end{figure}

In $\mathcal D_{\rm s}$ no trace of large scale patterning was found for $\Rn\ge410$ though disorganized regions where turbulence intensity was depleted, called {\it laminar troughs\/} in the following, were observed.
For $\Rn\le 395$, in all cases, the steady state regime obtained by continuing the simulation corresponded to one well-formed stable turbulent band either leaning to the left or to the right of the streamwise direction, as expected from symmetry considerations. 
The situation for $\Rn=400$ and $405$ was more complex, as can be understood from figure~\ref{Fet} which displays time-series of $E_{\rm t}$ as functions of time.
A large turbulent fraction implies a high value of $E_{\rm t}$, while the presence of laminar troughs corresponds to a lower value.
Parallel examination of snapshots taken all along the simulations helped me to characterize the flow regime.
In the examples shown, at $\Rn=405$ featureless turbulence is observed during most of the time except for a brief episode with a well-formed band.
At $\Rn=400$, featureless episodes are now scarce, separating longer periods where a definite pattern is conspicuous, i.e. left or right oriented single bands possibly perturbed by the presence of dislocations.
Reentrance of featureless turbulence at $\Rn=400$ and band intermittency at $\Rn=405$ are complementary facets that suggest one to locate $\Rt$ somewhere in-between and point to a small hysteresis characteristic of a discontinuous transition (weakly first-order in the terminology of phase transitions) further subjected to a strong noise associated with the background turbulence.

Results equivalent to those in $\mathcal D_{\rm s}$ were obtained in $\mathcal D_{\rm b}$ for $\Rn$ down to $410$ and, again, the neighborhood of $\Rt$ was crossed fast so that bands appeared for $\Rn\le 395$.
A well-formed three-band pattern developed from the states left at the end of the steps at $\Rn=395$ and $390$, whereas two bands emerged for $\Rn\le 380$.
In another experiment, by decreasing $\Rn$ more carefully by steps of length $\delta t=3500$ still with $\delta\!\Rn=5$ from a three-band pattern at $\Rn=380$, the three bands could be maintained down to $\Rn=340$.
At $\Rn=335$, a wavelength was lost in the spanwise direction {\it via\/} an Eckhaus-like instability and a two-band pattern was obtained that finally decayed for $\Rn<325$.
Since in the smaller system, the band pattern was observed to decay at $\Rn=330$ after a very long transient ($> 10^4\>h/U$), the two-band states obtained at $\Rn=330$ and $325$ are probably also transient, though decay was not observed before the programmed end of the experiment ($t_{\rm f}=5\times 10^3\>h/U$).

\begin{figure}
\begin{center}
\includegraphics[width=0.8\textwidth,clip]{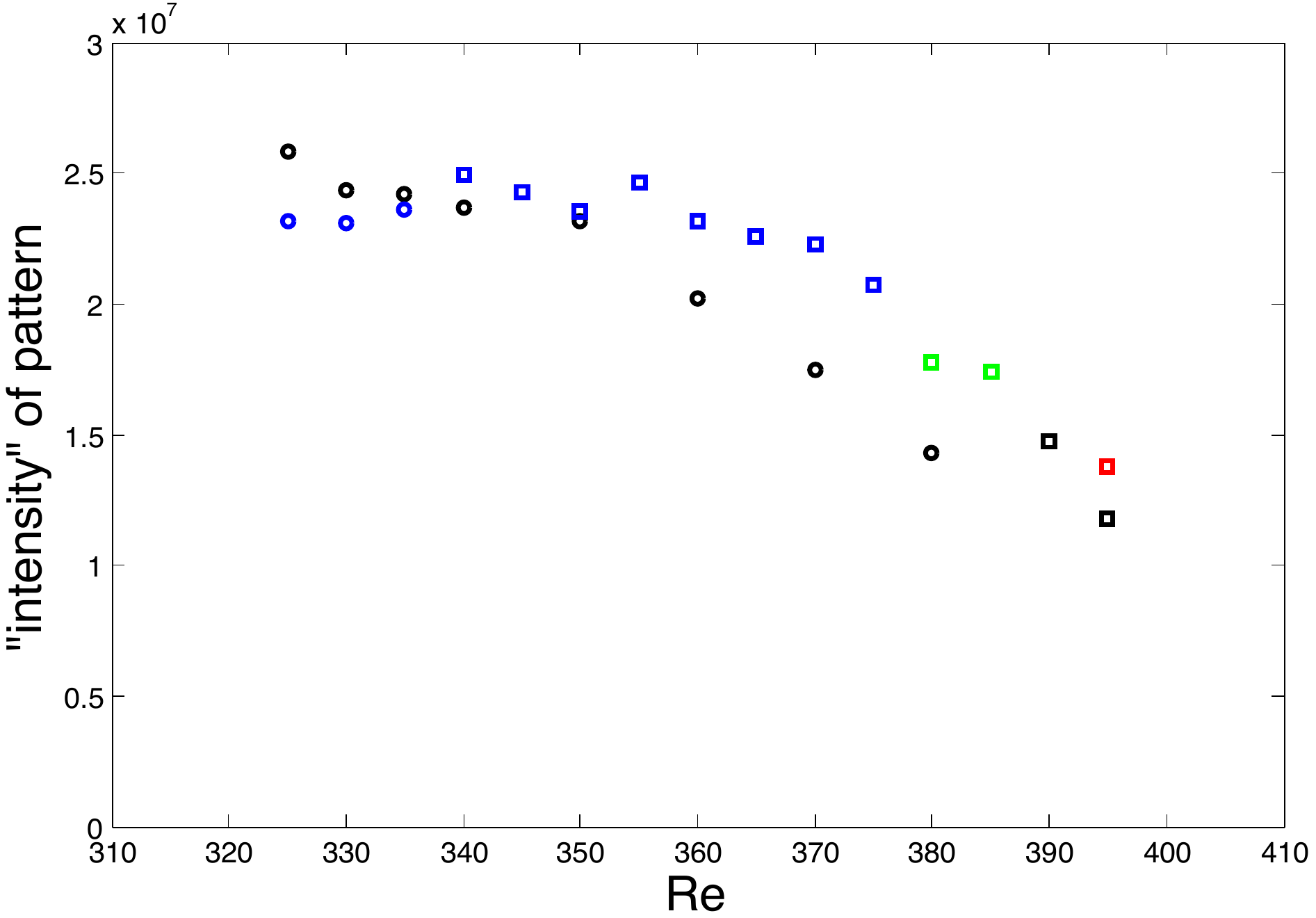}
\end{center}
\caption{\label{Fbdsp}Order in the pattern measured by the mean value of the modulus of the dominant Fourier component squared, here called `intensity', as a function of the Reynolds number for the large domain ($128\times160$). See explanations in the text.}
\end{figure}

The three-band state at $\Rn=380$ alluded to above was obtained from a two-band state at $\Rn=370$ that was Eckhaus-unstable upon increasing~$\Rn$.
So, in $\mathcal D_{\rm b}$, there exists a whole range of multi-stability with hysteresis, where two- and three-band patterns can be obtained depending on the experimental protocol.
This is summarized in figure~\ref{Fbdsp} that displays the intensity of patterning as a function of $\Rn$,
i.e. the squared modulus of  the dominant peak in the central part of the spatial Fourier spectrum of $e_{\rm t}(x,z,t)$ averaged over at least 80 spectra separated by at least 50 $h/U$  (units for turbulence intensity are arbitrary).
Standard deviations have been computed but not shown here for the sake of clarity because points all correspond to well-established patterns, the intensity of which does not fluctuate much.

Points shown correspond to spectra displaying a {\it single\/} dominant peak, either two bands, i.e.  with wave-numbers $(1,+2)$ or $(1,-2)$, or three bands, i.e. $(1,+3)$ or $(1,-3)$.
Their respective harmonics $(2,\pm4)$ or $(2,\pm6)$ were about one order of magnitude smaller, significantly above the level of the rest of the spectrum.
Squares and circles correspond to modes with wave-numbers $(1,\pm3)$ and $(1,\pm2)$, respectively. Different colors (on line;  in print, various shades of gray) correspond to experiments performed under different protocols.
The experiment with $\delta t=500$ is shown in black and the one with  $\delta t=3500$ in blue.
The two marks in green at $\Rn=380$ and $385$ have been obtained from the corresponding result at $\Rn=370$ and the mark in red at $\Rn=395$ from the three-band state at $\Rn=390$.

As noted earlier, only values corresponding to simple final states considered as steady over sufficiently long (but finite) durations are reported, hence nothing for  $\Rn\ge400$.
Multi-stability is easily understood as a result of periodic boundary conditions.
Discrete jumps from one orientation to the other are therefore similar to those observed in CCF~\cite{Pr01,Petal02} where they were due to periodicity along the azimuthal direction.
The more continuous variation experimentally observed in PCF~\cite{Pr01,Petal02} is presumably due to less stringent streamwise and spanwise boundary conditions.
\begin{figure}
\begin{center}
\includegraphics[width=0.4\textwidth,clip]{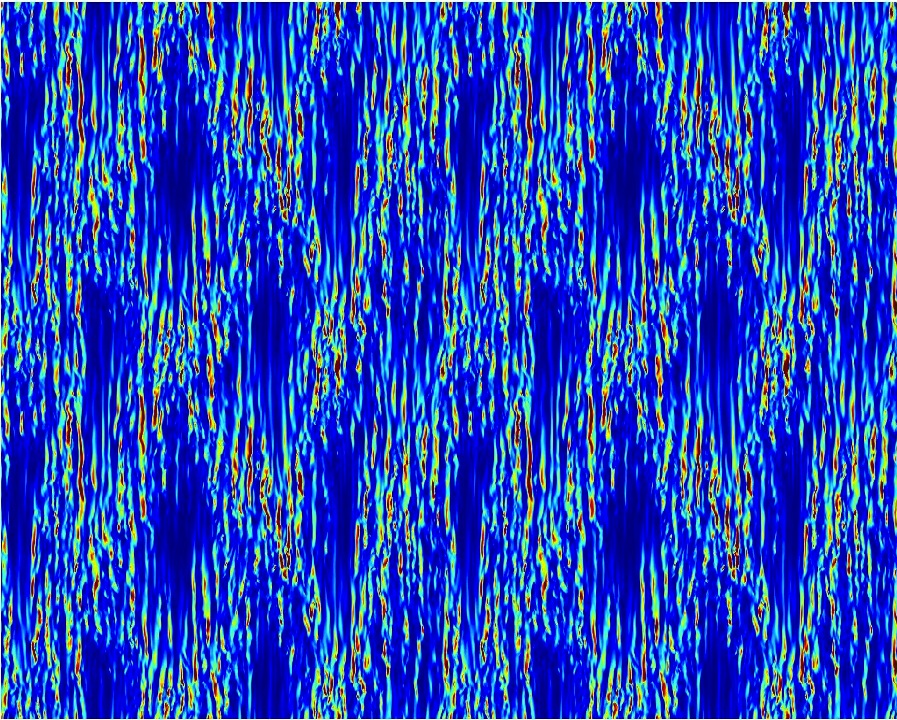}\hskip1em
\includegraphics[width=0.4\textwidth,clip]{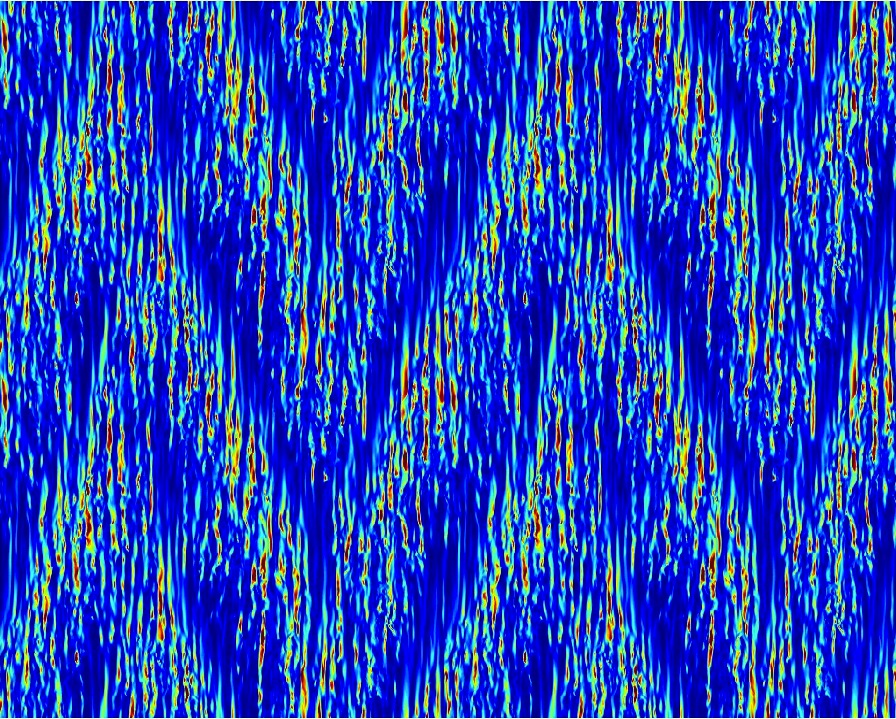}\\[2ex]
\includegraphics[width=0.4\textwidth,clip]{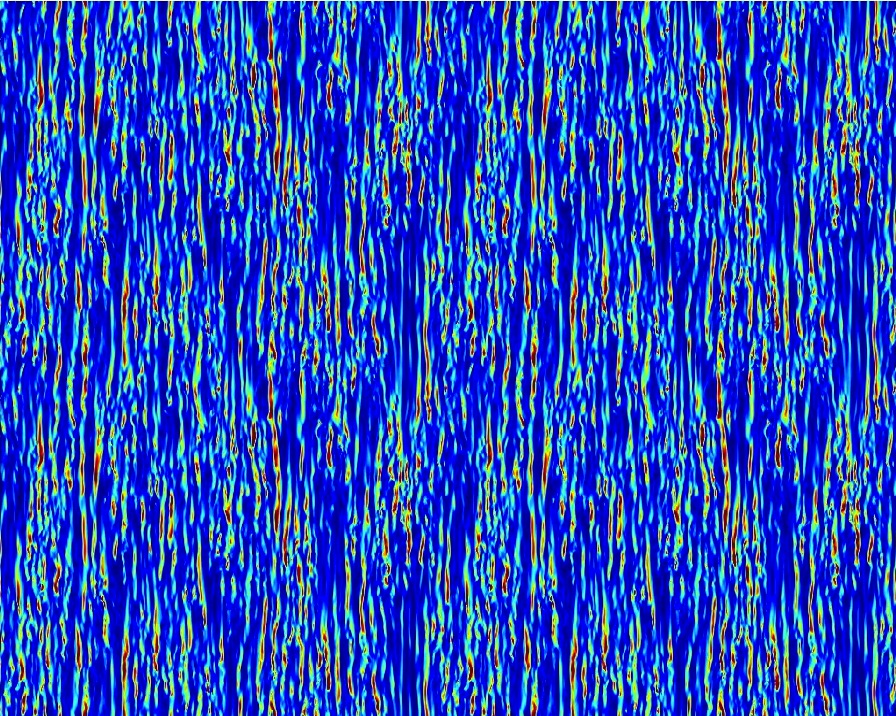}\hskip1em
\includegraphics[width=0.4\textwidth,clip]{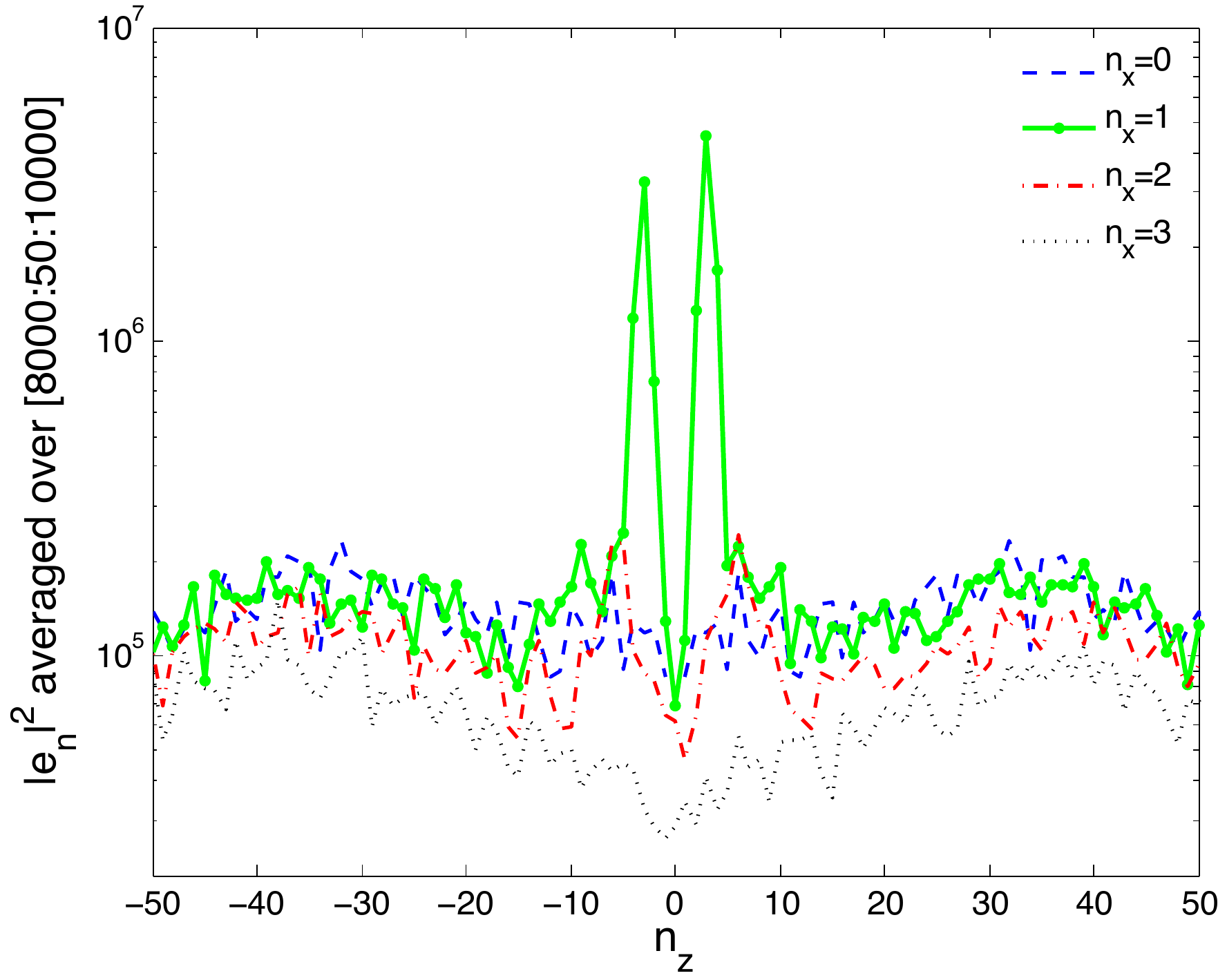}

\end{center}
\caption{\label{Fr400}Snapshots of the solution in an experiment at $\Rn=400$ for $t=7450$ (top left), $9600$ (top right), and $11800$ (bottom left), featuring the turbulent energy averaged over the upper layer $0\le y\le1$ (color levels on line).
Streamwise direction is vertical.
In order to enhance the appearance of the patterns, the images display a $2\times2$ tiling of the pattern permitted by the in-plane periodic boundary conditions.
Bottom-right: Center of the Fourier spectrum in the chevron pattern displayed just above.}
\end{figure}

No data can be seen in the upper part of the transitional range for $\Rn>\nobreak395$, the center of the spectrum was always found with complex structure forbidding the unambiguous isolation of a dominant mode.
As illustrated in figure~\ref{Fr400}, for $\Rn=400$, intermittent staggered arrangements of streamwise elongated laminar troughs ($t=7450$) evolved into chevon patterns ($t=9600$) that later decayed to mostly featureless turbulent flow riddled with smaller laminar patches ($t=11800$). 
The Fourier spectrum averaged over the period when the chevrons are conspicuous (bottom right panel) accordingly shows two symmetrical peaks but, in contrast to what happens at lower $\Rn$, they are broadened and conditional averaging has to be performed in order to obtain them.
Only the central part $n_x\le3$, $|n_z|\le 50$ is shown; the broad humps centered around $|n_z|=35$ corresponds to the streamwise streaks with mean spanwise period $\approx L_z/35=4.6$.
The chevrons are reminiscent of what was observed in the upper transitional range of CCF~\cite{Petal02}, except that here they are intermittent rather than steady.
\begin{figure}
\begin{center}

\includegraphics[width=0.4\textwidth,clip]{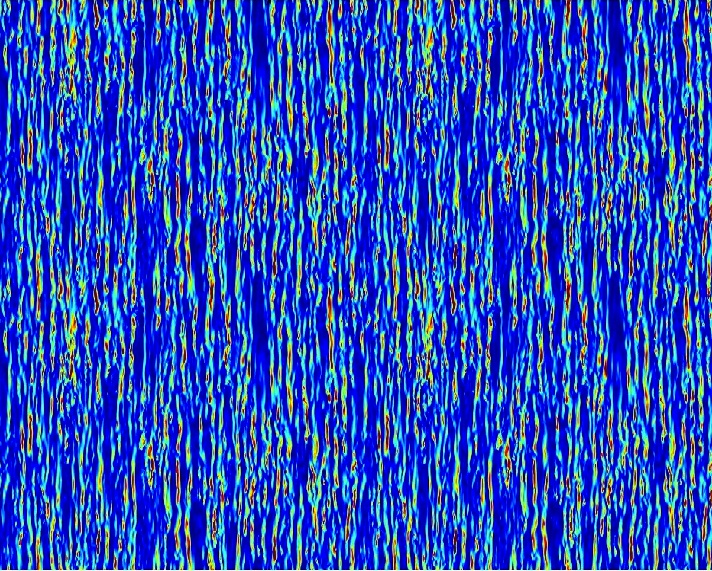}\hskip1em
\includegraphics[width=0.4\textwidth,clip]{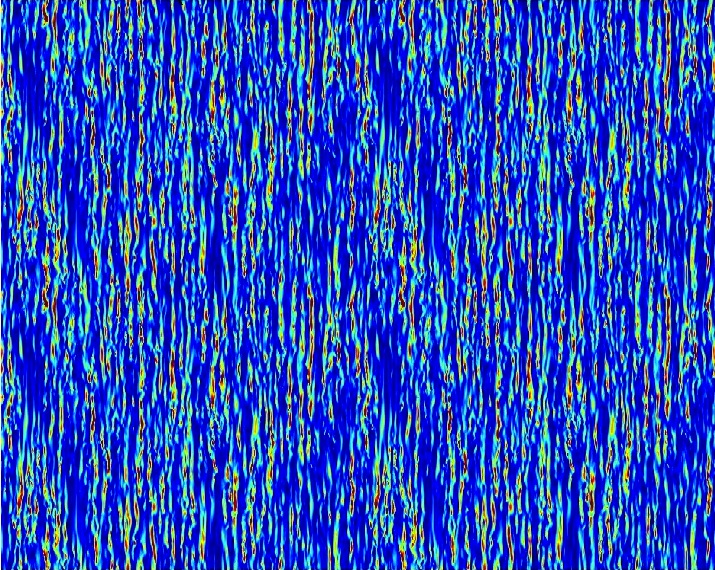}\\[2ex]
\includegraphics[width=0.4\textwidth,clip]{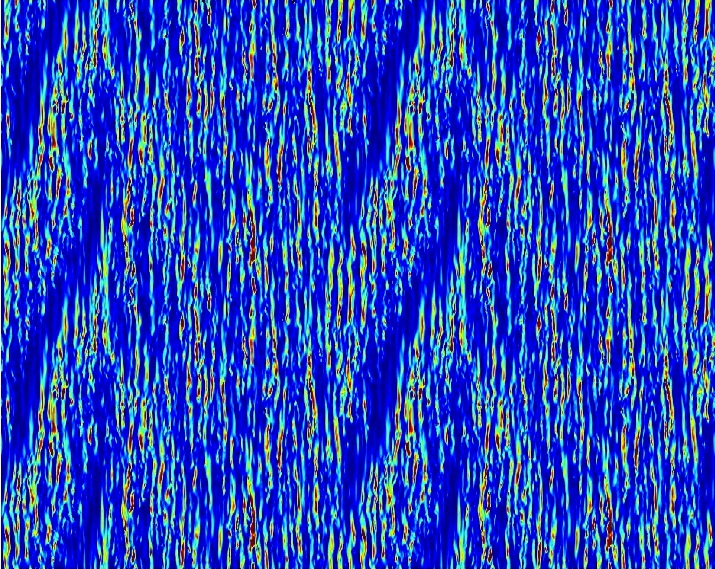}\hskip1em
\includegraphics[width=0.4\textwidth,clip]{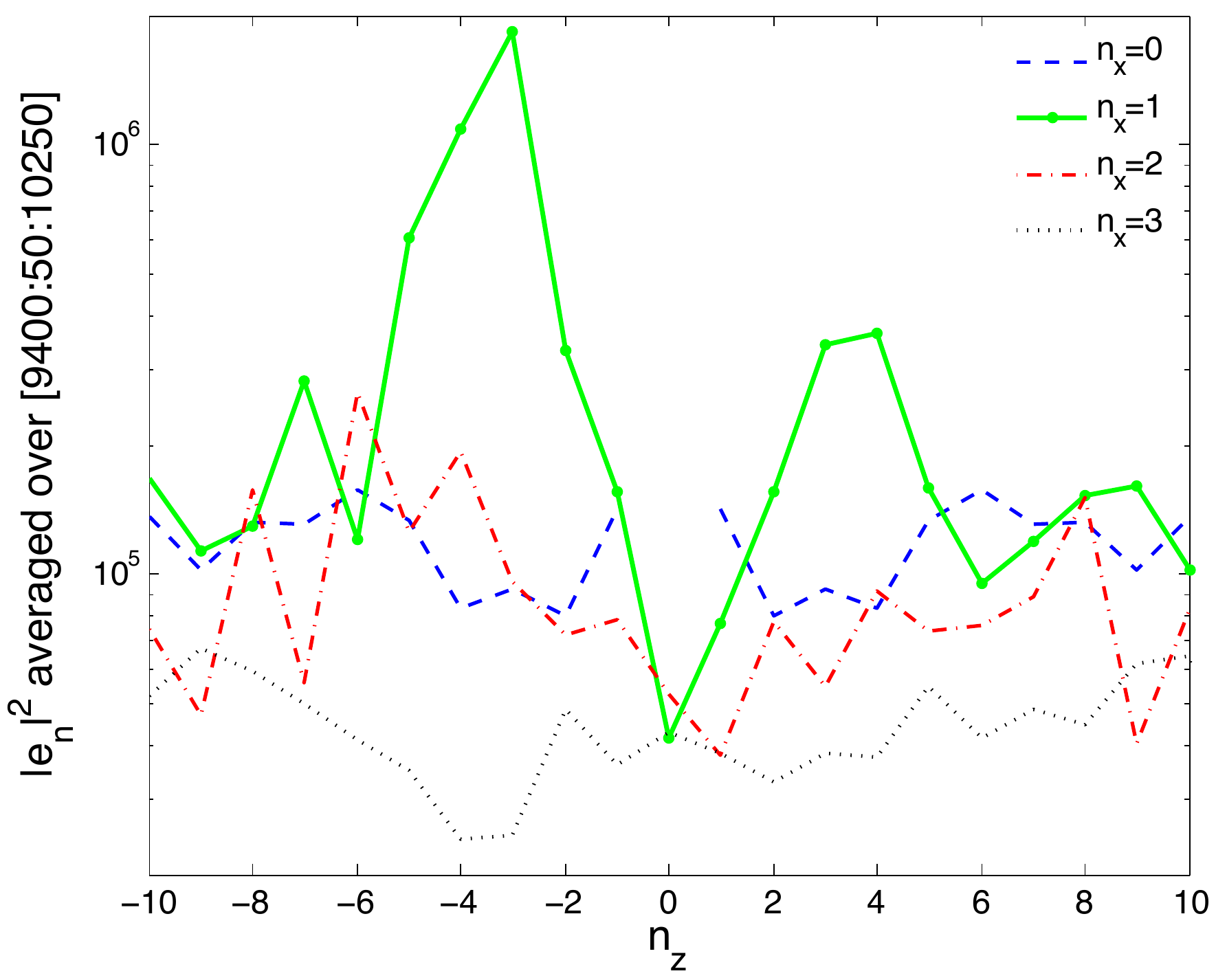}

\end{center}
\caption{\label{Fr405}Snapshots of the solution in an experiment at $\Rn=405$ for $t=7350$ (top-left), $t=8100$ (top-right), and $10050$ (bottom left); same representation as for images in figure~\ref{Fr400}. Bottom-right: Center of the Fourier spectrum corresponding to localized oblique laminar ribbons on the left.}
\end{figure}

At larger $\Rn$, patterns organized over the whole width of the domain have not been observed but only localized tatters appearing intermittently, either streamwise elongated patches or oblique laminar ribbons as illustrated in figure~\ref{Fr405} for $\Rn=405$. 
The bottom-right panel displays the very center of the Fourier spectrum corresponding to the localized oblique laminar ribbons shown in the bottom-left panel. These ribbons make an angle of $20.2^\circ$ with respect to the streamwise direction, i.e. less than $22.6^\circ$ for a three-band pattern fitting the domain but more than $17.4^\circ$ for an ideal four-band pattern.
The angle and spanwise localization manifest themselves in the Fourier spectrum that, besides a main peak at $n_z=-3$, contains substantial power at $n_z=-4$ and also significant power at $n_z=-2$ and $-5$, as well as in a wing at positive $n_z$ from $+2$ to $+5$, which nicely account for a main mode modulated by a localized envelope.

For $\Rn\ge405$, oblique patches become rare but streamwise elongated ones persist  to large values of $\Rn$ though becoming smaller and smaller.
Fourier spectra conditionally averaged over short durations may accidentally present some structure reflecting the short lifetime of laminar troughs arrangements.
However, above $\Rn\simeq400$, the determination of an order parameter in the sense of phase transitions from Fourier spectra seems quite delicate.
Progress in this domain (larger domains approaching the ``thermodynamic limit'' and longer durations for better statistics) would be obtained only through a detailed study of space-time correlations that is much beyond my present computational capabilities.

\subsection{Discussion\label{A1.3}}

These results compare favorably with earlier empirical  results, experimental~\cite{Petal02} and numerical~\cite{Detal10b}, giving a reliable quantitative access to the orientation changes with $\Rn$.
In particular, images in figures~\ref{Fr400} and \ref{Fr405} are remarkably similar to those in figure~3.28 in Prigent's thesis~\cite{Pr01} at similar Reynolds numbers (see note~\ref{FN4}).
The periodic boundary conditions however freeze the orientation over finite intervals of $\Rn$, implying slightly subcritical transitions {\it via\/} Eckhaus-like dislocation nucleation ($3\to2$ bands at $\Rn\gtrsim335$, $2\to3$ bands at $\Rn\lesssim380$) and the intermittent occurrence of laminar troughs leaning at smaller angles for $\Rn\gtrsim400$.
They also contradict observations made in~\cite{Tetal09} where a continuous transition at a much higher threshold $\Rt\sim 440$ was mentioned.
This discrepancy can be attributed to the role of the geometry used by Barkley and Tuckerman which, forbidding orientation fluctuations and severely reinforcing the streamwise coherence, has a tendency to delay the occurrence of featureless turbulence towards larger~$\Rn$, whereas the quasi-1d character of the domain justifies the observed continuous nature of the transition as in other similar circumstances~\cite{CM95,Ba11,Setal13}.

\section{About models of laminar-turbulent patterning\label{A2}}

\subsection{Context\label{A2.1}}

Much as Waleffe's modeling~\cite{Wa95} accounts for the process by which turbulence is  locally maintained, the modeling adapted to extended systems is expected to enlighten the nature of the laminar-turbulent coexistence characterizing the subcritical transition to turbulence in wall-bounded flows.

The pioneering work of Barkley for pipe flow~\cite{Ba11,Ba11b} treats turbulence as the result of a local ``chemical reaction'' between two variables, one characterizing the mean shear ($u$) and the other the local turbulence level ($q$).
Spatial dependence along the axis of the pipe is next introduced in the form of an effective diffusion of $q$.
Adding axial transport of $u$ and $q$ then leads to what is known as an {\it advection-reaction-diffusion} model.
A clever choice of the reaction part led him to account for the whole transitional range quantitatively by scaling $\Rn$ appropriately.
An {\it ad hoc\/} extension to PCF~\cite{Ba11b} was however less convincing.
Keeping up with the reaction-diffusion concept, in ref.~\cite{Ma12} I proposed to interpret patterning as the result of a Turing instability~\cite{Mu93}.

The model supporting my conjecture relied on Waleffe's system~\cite{Wa97} for the local reaction part. 
The corresponding dynamical system, which does not encrypts any large-scale space dependence, can be termed {\it zero-dimensional\/} (0d).
This assumption was relaxed by introducing, on purely phenomenological grounds, effective diffusivities akin to turbulent viscosities in the direction of the wavevector of the putative pattern, i.e. a {\it one-dimensional\/} (1d) reaction-diffusion model.
According to the mechanism of a standard Turing instability~\cite{Mu93}, when the diffusion coefficient controlling streamwise perturbations (Waleffe's variable $M$, similar to Barkley's $u$) was taken much larger than that of the streak instability mode (variable $W$, playing the role of $q$) a periodic pattern was obtained, retaining the subcritical character of the transition at $\Rg$~\cite{Ma12}.
Despite its interest, the main limitations of the model were the simplicity of the large-scale dynamics reduced to a naive but plausible turbulent diffusion and, above all,  the arbitrariness in the orientation of the space coordinate unable to explain the origin of the obliqueness of the laminar-turbulent bands, both calling for a more realistic introduction of large scale dependence in physical space. 

My involvement in the construction of models based on {\it weigthed residual\/} approximations~\cite{Fi72} of primitive equations dates back to the early {\it Eighties}, at that time for thermal convection~\cite{Ma83}. 
Recently, I attacked the problem of 2d modeling of PCF in the same perspective.
The derivation first involves  a systematic decomposition of the wall-normal  ($y$) dependence of the hydrodynamic fields on a complete functional basis.
Next, a projection of the full three-dimensional Navier--Stokes equations is performed on the same basis, which singles out the Galerkin method among the vast family of weighted residual methods~\cite{Fi72}.
This leaves one with an infinite set of partial differential equations depending on the in-plane coordinates ($x,z$) and time~$t$.
Finally, a severe truncation of this set is performed by retaining the very first Galerkin amplitudes, which yields a 2D model partially accounting for the third dimension through the shape of the corresponding wall-normal modes.
 
For PCF, this approach was first developed using the trigonometric basis permitted by stress-free boundary conditions \cite{MD01}.
Yielding the most natural 2D extension of Waleffe's model, it reproduced the local SSP and the subcritical nature of the transition, but strongly underestimated $\Rg$.
The model was subsequently adapted to no-slip boundary conditions with M.~Lagha \cite{LM07a} using the Galerkin basis introduced in~\cite[p.199ff]{GP71} that I previously considered in the context of thermal convection~\cite{Ma83}.
A more realistic $\Rg$ was obtained but  the expected band pattern failed to show up  \cite{Ma09}.
This failure was attributed to an exaggeratedly reduced effective wall-normal resolution since direct numerical simulations have shown that a slight improvement is sufficient to recover the bands \cite{MR11}.
A straightforward option was therefore to push the truncation of the Galerkin expansion to a higher order.

\subsection{A model for plane Couette flow with some preliminary results\label{A2.2}}

The improved model derived with K.~Seshasayanan~\cite{Se13,MS14} was obtained by closely following the approach developed with M.~Lagha~\cite{LM07a}.
The main novelty was that, to avoid difficulties linked to the treatment of the continuity equation, we adopted a velocity-vorticity formulation  ($v,\zeta)$ \cite[p.155ff]{SH01} but we used the same basis.
The wall-normal velocity and vorticity components were expanded as
$$
v=\sum_{j=1}^{j_{\rm max}} V_j(x,z,t) f_j(y)\,, \qquad\zeta\equiv\partial_z u -\partial_x w=\sum_{j'=0}^{j'_{\rm max}} Z_{j'}(x,z,t) g_{j'}(y)\,,
$$
with $f_j(y)=(1-y^2)^2 P_j(y)$ and $g_{j'}(y)=(1-y^2) Q_{j'}(y)$. Pre-factors $(1-y^2)^2$ and $(1-y ^2)$ ensure that the boundary conditions on $v$ and $\zeta$ at $y=\pm1$, $v=\partial_y v=\zeta=0$, are automatically fulfilled.
 Polynomials $P_j(y)$ and $Q_{j'}(y)$ are of degree $j-1$ and $j'$.
This labeling is chosen so that, when taking the continuity equation into account, a consistent truncation is obtained with $j_{\rm max}=j'_{\rm max}$, hence three fields for $j_{\rm max}=1$ in \cite{LM07a}.

The generic form of the equations for amplitudes $V_j$ and $Z_j$, given in~\cite{MS14} which is currently not accessible, is repeated here for more specificity:
\begin{eqnarray*}
&&\left[\left(\delta^i_j(\partial_{xx}+\partial_{zz}) +a^i_j\right)\partial_t + \left({\delta'}^i_j(\partial_{xx}+\partial_{zz}) + {a'}^i_j\right)\partial_x\right] V_i + N^{(V)}_j\\
&&\hspace{8em}=\mathrm{Re}^{-1}\left[ \delta^i_j (\partial_{xx}+\partial_{zz})^2 + 2 k^i_j (\partial_{xx}+\partial_{zz}) + m^i_j\right] V_i\,,\\
&&\left(\delta^i_j \partial_t + b^i_j \partial _x\right) Z_i + {b'}^i_j \partial_z V_i + N^{(Z)}_j= \mathrm{Re}^{-1} \left[ \delta^i_j(\partial_{xx}+\partial_{zz}) + n^i_j\right]Z_i\,.
\end{eqnarray*}
Nonlinear terms $N^{(V)}_j$ and $N^{(Z)}_j$ are quadratic expressions of the  velocity component amplitudes $U_j$, $V_j$, $W_j\>$:
\begin{eqnarray*}
&&N^{(V)}_j=(\partial_{xx}+\partial_{zz}) \left\{q_j^{ik}\left[\partial_x(U_iV_k)+\partial_z(W_iV_k)\right]+{q'}_j^{ik}(V_i V_k)\right\}\\
&&\hspace{3.5em}\mbox{}-\,r_j^{ik}\left[\partial_{xx} \left(U_iU_k\right) + 2\partial_{xz}\left(U_iW_k\right)+\partial_{zz} \left(W_iW_k\right)\right]\\
&&\hspace{3.5em}\mbox{}-{r'}_j^{ik}\left[\partial_x\left(U_iV_k\right)+\partial_z\left(W_iV_k\right)\right],\\ 
&&N^{(Z)}_j= s^{ik}_j\left[\partial_{xz} (U_iU_k - W_iW_k) + (\partial_{zz}-\partial_{xx})(U_jW_k)\right]\\
&&\hspace{3.5em}\mbox{}+{s'}^{ik}_j \left[\partial_z(U_iV_k) -\partial_x(W_iV_k))\right].
\end{eqnarray*}
The in-plane velocity components contain uniform contributions that have to be dealt with separately~\cite{SH01}.
Setting $U_j=\overline U_j+\widetilde U_j$, $W_j=\overline W_j+\widetilde W_j$, for the uniform parts $(\overline U_j, \overline W_j)$ one gets:
$$
\delta^i_j\partial_t \overline U_i  + {s'}^{ik}_j\,\overline{U_iV_k}=\mathrm{Re}^{-1}\, n^i_j\, \overline U_i \,,\qquad
\delta^i_j\partial_t \overline W_i + {s'}^{ik}_j\, \overline{W_iV_k}=\mathrm{Re}^{-1}\, n^i_j\, \overline W_i \,,
$$
where the overline means space averaging over the whole domain under study (in-plane periodic boundary conditions assumed).
Everywhere repeated  identical upper and lower indices correspond to implicit summation, all coefficients being straightforwardly obtained as integrals of polynomials over $y\in[-1,+1]$ at the projection step.

Besides being easy to use the chosen basis captures the wall-normal dependence of the flow quite efficiently, as illustrated in figure~\ref{Fcomp}.
Two reference numerical solutions obtained in the large domain ($128\times160$) of Appendix~A (subscript `ref' in the following) have been projected on the polynomial basis and then reconstructed using an increasingly large number of Galerkin amplitudes for $\Rn=370$ (two band solution) and $\Rn=450$ (featureless state).
The left panel in figure~\ref{Fcomp} displays the perturbation energy contained in the different reconstructions compared to that in the reference states.
In the right panel, the nearly exponential decrease of the point-wise residual
$\int (\mathbf v-\mathbf v_{\rm ref})^2 
/\int (\mathbf v_{\rm ref})^2 
$ gives a good idea of the convergence of the expansion.
The reconstruction is marginally better at $\Rn=370$ than at $\Rn=450$.
This can be understood from the contribution of the laminar region which is certainly better rendered than the turbulent region that, whether banded or featureless, requires better resolution close to the walls. 
\begin{figure}
\begin{center}

\includegraphics[height=0.34\textwidth,clip]{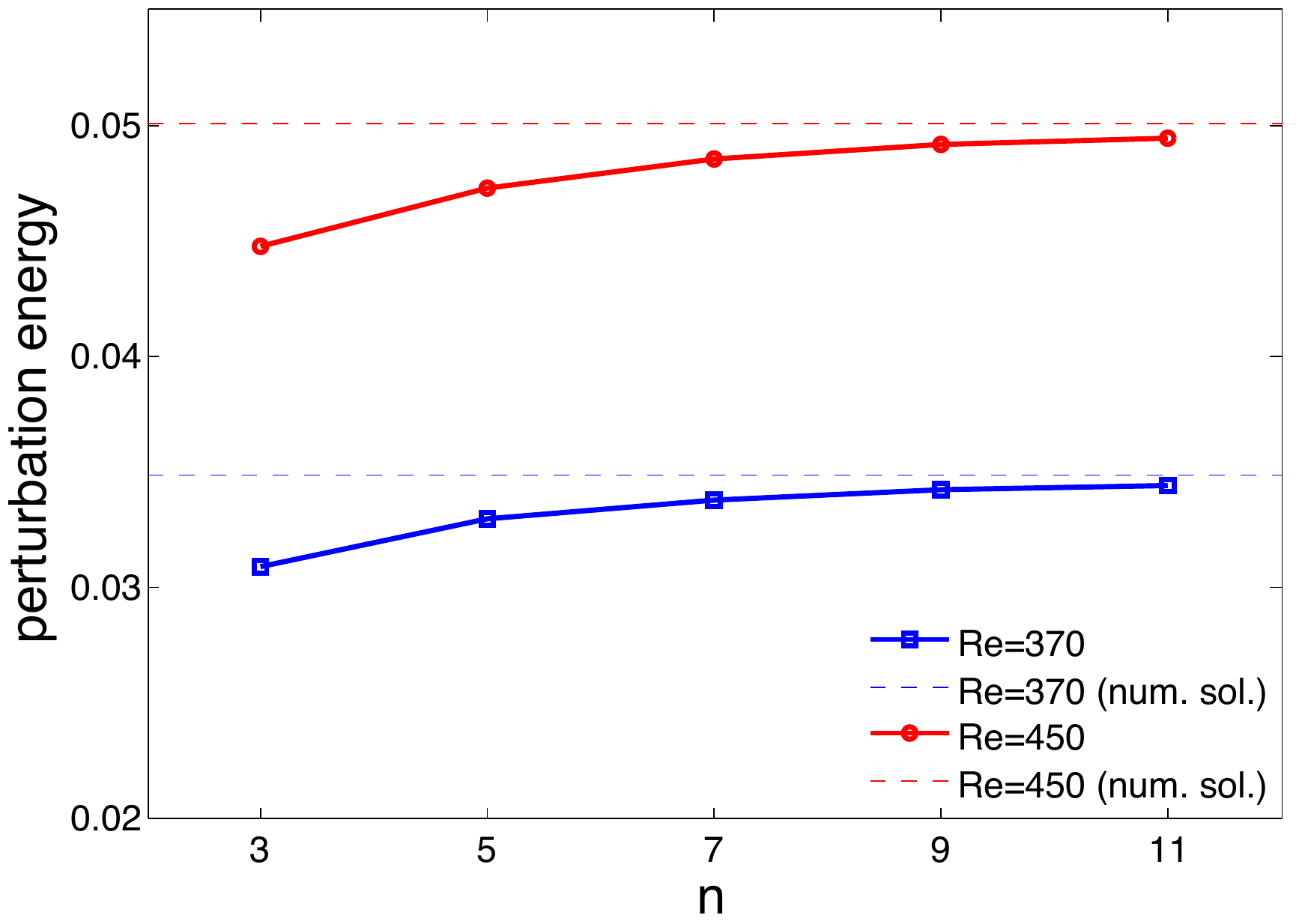}\hskip0.05\textwidth
\includegraphics[height=0.34\textwidth,clip]{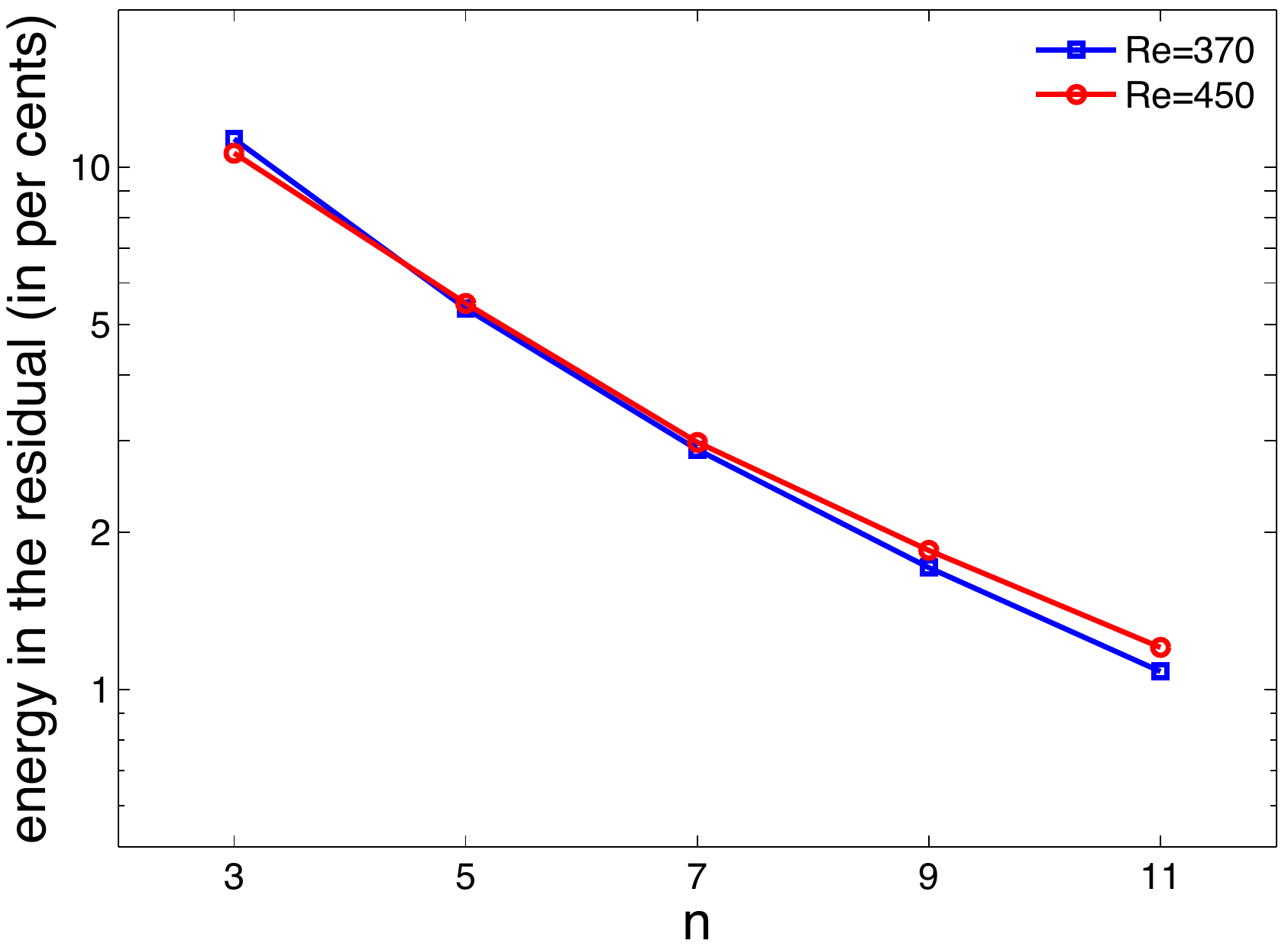}

\end{center}
\caption{\label{Fcomp} Left: Perturbation energy contained in the reconstructed solution as a function of the number of polynomials compared to that in the original numerical solution (dashed line). Right: Relative residual energy. $\Rn=370$ as blue square dots, and $\Rn=450$ as red round dots. (Color on line).}
\end{figure}

\begin{figure}
\begin{center}

\includegraphics[width=0.49\textwidth,clip]{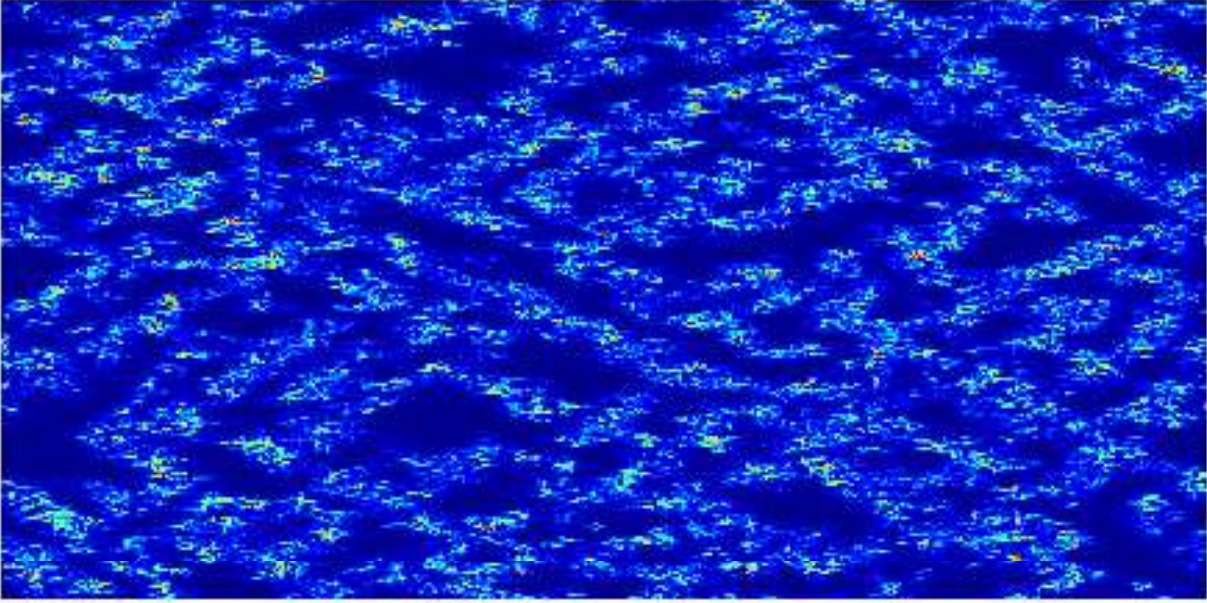}
\includegraphics[width=0.49\textwidth,clip]{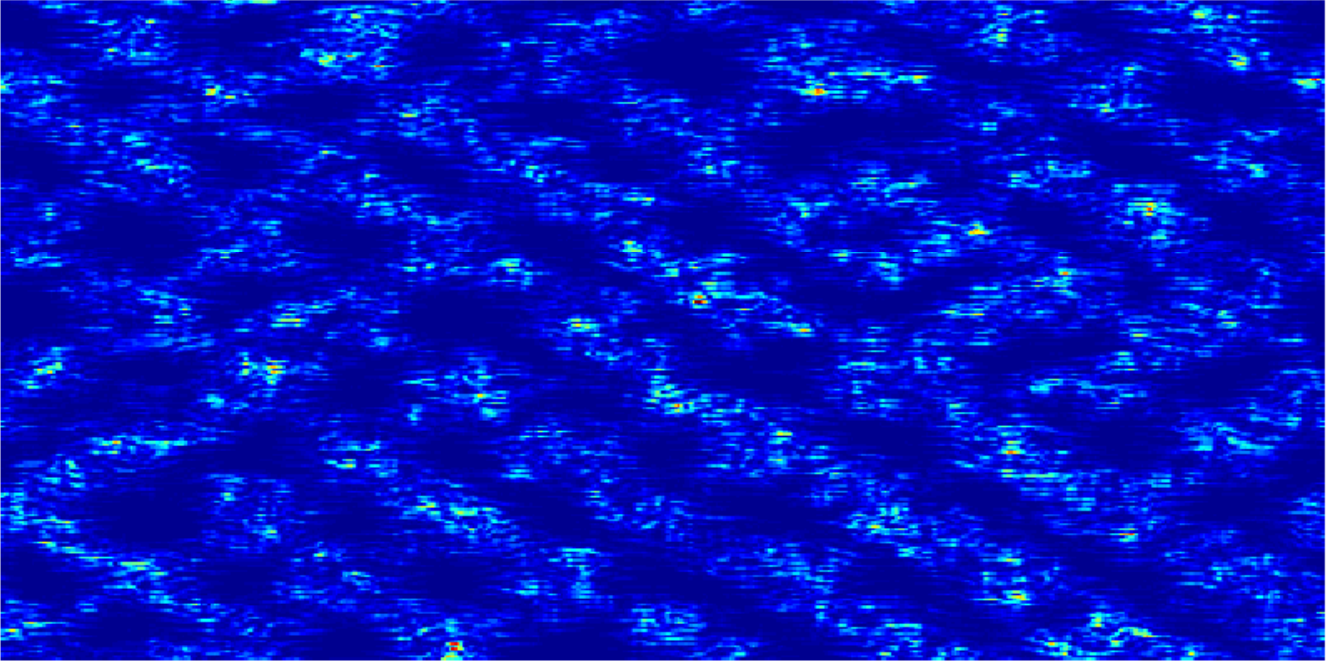}

\end{center}
\caption{\label{F7fm} Field of perturbation energy from a simulation of the seven-field model at $\Rn=151$ (left, courtesy of K.~Seshasayanan) and from a DNS using {\sc ChannelFlow}~\cite{Gix} with $N_y=11$ Chebyshev polynomials at $\Rn=250$, adapted from~\cite{MR11}. Domain size is $\simeq 680\times340$. (Color levels on line).}
\end{figure}

The numerical implementation by K.~Seshasayanan and some preliminary simulations are described in detail elsewhere~\cite{Se13}.
Truncation at $j_{\rm max}=3$, i.e. seven Galerkin amplitudes, has been selected because it allowed us to include one correction of each parity, odd and even, to the previous three-field model of~\cite{LM07a} while being presumably accurate enough (figure~\ref{Fcomp}, right panel).
Figure~\ref{F7fm} (left panel) displays a snapshot of the perturbation energy contained in the solution at steady state for $\Rn=151$ in a domain of size $680\times340$, comparable to that of Prigent's experimental setup~\cite{Pr01}.
From that picture it is clear that the model is able to reproduce the patterning in its own transitional range $[\Rg,\Rt]\sim[150,159]$.
The right panel of figure~\ref{F7fm} shows the state of the flow in a domain of nearly identical size ($682\times341$) obtained using {\sc ChannelFlow} like in~Appendix~A but strongly under-resolved in wall-normal direction with $N_y=11$ Chebyshev polynomials~\cite{MR11} instead of 33 above.

In both cases, patterning appears at values of $\Rn$ smaller than in experiments~\cite{Petal02} or in the well-resolved numerics of \cite{Detal10b} and Appendix~A, for the reason already indicated in \S\ref{S1.2} about under-resolved DNSs.
Of course this effect is even more important in the seven-field model than is the DNS with $N_y=11$ that resolves the flow more finely though not yet enough to recover the actual transitional range.
From figure~\ref{F7fm}, it is also apparent that the pattern obtained with the model has smaller streamwise and spanwise wavelengths than expected.
This is again an effect of effective wall-normal under-resolution since DNSs with Chebyshev polynomials show that, when $N_y$ is decreased, the streamwise coherence of streaks is diminished so that for $N_y<11$ remaining turbulent fields do not show bands and resemble those obtained with the three-field model~\cite{MR11}.
In this respect, polynomials used here seem to do better since the streamwise coherence is indeed impaired but bands with physically credible obliquity are still present. 

\subsection{Perspectives\label{A2.3}}

Contrasting with under-resolved simulations that behave as black boxes, the model presents itself as a closed set of equations.
It has been obtained in a systematic way and seems to encode the most relevant features of the transition to turbulence in PCF including the bands, provided that truncation takes place just beyond lowest order, i.e. seven fields.
Going further with analytical approximations is thus manageable {\it via\/} explicit scale separation (local SSP {\it vs.}~pattern), averaging, and adiabatic elimination of enslaved degrees of freedom, yielding higher degree terms from quadratic ones as in~\cite{Ma12}, and explicit anisotropic expressions for the turbulent diffusion terms.
Remarkably enough, the complete solution contains uniform in-plane mean-flow components~\cite[p.155ff]{SH01} that are obtained as space averages of Reynolds stresses~\cite{MS14} responsible for a non-local feedback (see the corresponding equations in the model above).
This feedback term closely corresponds to the integral term controlling the relative width of the turbulent and laminar domains in cylindrical Couette spiral turbulence, as introduced by Hayot \& Pomeau in their  RGL5 model~\cite{HP94}.
An effective reduced model -- still being developed -- is thus expected to emerge from the whole procedure.
Beyond the 1d model that I introduced in~\cite{Ma12}, it should thus be fully realistic and, wishfully, able to predict (and explain the origin of) the patterning in PCF and other similar flows.

\paragraph{Acknowledgments} It was my pleasure to accept the invitation of Lutz Lesshaft to contribute to this special issue celebrating Patrick Huerre.
My acquaintance with Patrick dates back to the mythical {\it Sixties\/} and was boosted by my joining LadHyX soon after the lab's foundation.
We should all thank him for  the quality of the work he knew how to foster and for the cool atmosphere he managed to maintain in the lab, both making the distinctive feature of LadHyX.
Now, I would like to express my gratitude to all the past and present collaborators who contributed to my understanding of the problem of the transition to turbulence either subcritical or supercritical, at GIT-Saclay, LadHyX, ENSTA, LIMSI, ESPCI, and elsewhere.
Fearing to forget one, I shall not name anybody except the most recent, K.~Seshayanan, who provided me with the simulation results presented in Appendix~B.
A special mention is however due to Yves Pomeau who introduced me to the arcana of the nonlinear world.
Patrick and the two referees were instrumental in improving the manuscript and, as such, are also deeply acknowledged.

\bigskip
\noindent{\bf Literature cited}

\end{document}